\documentclass[a4paper,fleqn,usenatbib]{mnras}

\usepackage{newtxtext}
\usepackage{newtxmath}

\usepackage{xspace}

\usepackage[T1]{fontenc}
\usepackage{ae,aecompl}
\usepackage[autostyle]{csquotes}

\usepackage{graphicx}	
\usepackage{amssymb}	

\usepackage{CJKutf8}

\newcommand{\nup}{\ensuremath{\nu}p\xspace}
\newcommand{\ye}{\ensuremath{Y_\mathrm{e}}\xspace}
\newcommand{\triplea}{3$\alpha$ reaction\xspace}
\newcommand{\iso}[2]{\ensuremath{{}^{#2}\mathrm{#1}}\xspace}
\newcommand{\reac}[5]{$\iso{#1}{#2}$(#3)$\iso{#4}{#5}$}
\newcommand{\npreac}[5]{$\iso{#4}{#5}$(n,\,p)$\iso{#1}{#2}$}

\title[Nucleosynthesis uncertainties in the \nup process]{Uncertainties in \nup-process nucleosynthesis from Monte Carlo variation of reaction rates}

\author[Nishimura~el~al.]
{
N.~Nishimura \begin{CJK}{UTF8}{ipxm}(西村信哉)\end{CJK}$^{1, 2}$\thanks{e-mail: nobuya.nishimura@yukawa.kyoto-u.ac.jp}\thanks{UK Network for Bridging Disciplines of Galactic Chemical Evolution (BRIDGCE), \url{https://www.bridgce.ac.uk}},
T.~Rauscher$^{3, 4}$\footnotemark[2],
R.~Hirschi$^{2, 5}$\footnotemark[2],
G.~Cescutti$^{4,6}$\footnotemark[2],\newauthor
A.~St.~J.~Murphy$^{7}$\footnotemark[2],
and C. Fr\"{o}hlich$^{8}$
\\
  $^1$ Center for Gravitational Physics, Yukawa Institute for Theoretical Physics, Kyoto University, Kyoto 606-8502, Japan\\
  $^2$ Astrophysics Group, Faculty of Natural Sciences, Keele University, Keele ST5 5BG, UK\\
  $^3$ Department of Physics, University of Basel, 4056 Basel, Switzerland\\
  $^4$ Centre for Astrophysics Research, University of Hertfordshire, Hatfield AL10 9AB, UK\\
  $^5$ Kavli IPMU (WPI), University of Tokyo, Kashiwa 277-8583, Japan\\
  $^6$ INAF, Osservatorio Astronomico di Trieste, I-34131 Trieste, Italy\\
  $^7$ School of Physics and Astronomy, University of Edinburgh, Edinburgh, EH9 3FD, UK\\
  $^8$ Department of Physics, North Carolina State University, Raleigh, NC 27695-8202, USA
}

\date{Accepted XXX. Received YYY; in original form ZZZ}

\pubyear{2019}

\begin{document}
\label{firstpage}
\pagerange{\pageref{firstpage}--\pageref{lastpage}}
\maketitle

\begin{abstract}
It has been suggested that a \nup process can occur when hot, dense, and proton-rich matter is expanding within a strong flux of anti-neutrinos. In such an environment, proton-rich nuclides can be produced in sequences of proton captures and (n,\,p) reactions, where the free neutrons are created in situ by $\overline{\nu}_\mathrm{e}+\mathrm{p} \rightarrow \mathrm{n}+\mathrm{e}^+$ reactions. The detailed hydrodynamic evolution determines where the nucleosynthesis path turns off from $N=Z$ line and how far up the nuclear chart it runs. In this work, the uncertainties on the final isotopic abundances stemming from uncertainties in the nuclear reaction rates were investigated in a large-scale Monte Carlo approach, simultaneously varying ten thousand reactions. A large range of model conditions was investigated because a definitive astrophysical site for the \nup process has not yet been identified. The present parameter study provides, for each model, identification of the key nuclear reactions dominating the uncertainty for a given nuclide abundance. As all rates appearing in the \nup process involve unstable nuclei, and thus only theoretical rates are available, the final abundance uncertainties are larger than those for nucleosynthesis processes closer to stability. Nevertheless, most uncertainties remain below a factor of three in trajectories with robust nucleosynthesis. More extreme conditions allow production of heavier nuclides but show larger uncertainties because of the accumulation of the uncertainties in many rates and because the termination of nucleosynthesis is not at equilibrium conditions. It is also found that the solar ratio of the abundances of \iso{Mo}{92} and \iso{Mo}{94} could be reproduced within uncertainties.
\end{abstract}




\section{Introduction}
\label{sec:intro}

The \nup process has been proposed to occur when hot, dense, and proton-rich matter is ejected from an astrophysical site under the influence of a strong neutrino flux. Such ejection can be found, e.g., in the dynamical ejecta of core-collapse supernovae (ccSNe) \citep{fro06a,fro06b}, in neutrino-driven proto-neutron-star (PNS) winds \citep{pruet06, 2006ApJ...647.1323W, 2011ApJ...729...46W}, in outflows from the massive PNS in ``hypernovae'' \citep{2015ApJ...810..115F}, and in outflows from collapsar models \citep{2010PhRvC..81b5802K}. Which sites actually experience a \nup process still partially remains an open question, the answer to which depends on the detailed hydrodynamic modeling of the outflows and the neutrino emission.

Regardless of the astrophysical site, the general features of the \nup process mainly depend on nuclear properties, such as reaction $Q$-values and reaction rates. They are briefly described below and in more detail in Section~\ref{sec:features}. In a \nup process, starting at \iso{Ni}{56}, sequences of proton captures and (n,\,p) reactions produce nuclei with larger and larger charge numbers $Z$ and mass numbers $A$ \citep{fro06a,fro06b,pruet06,2006ApJ...647.1323W}. During most of the nucleosynthesis timescale, proton captures and ($\gamma$,\,p) reactions are in equilibrium, similarly to an {\it rp} process, and the nucleosynthesis path up to Mo follows the $N=Z$ line in the nuclear chart (see Section~\ref{sec:features} for further discussion of the location of the \nup-process path). Below 1.5 GK, however, charged particle reactions freeze out quickly, leaving only (n,\,p) and (n,\,$\gamma$) reactions acting at late time which push the matter back to stability. After all other reactions have ceased, all remaining unstable nuclides decay to stability through electron captures or $\beta^+$ decays.

The amount of nuclei produced in the \nup process is small compared to that in the $s$ or $r$ process. Nevertheless, the \nup process may contribute to abundances not dominated by the $s$ and $r$ processes. This may be of relevance to explain high abundance ratios of Sr, Y, Zr relative to Ba in metal-poor stars \citep{2007A&A...476..935F,2007ApJ...671.1685M,2014JPhG...41d4005A}. The \nup process could also provide an important contribution to the lighter $p$ nuclides\footnote{Proton-rich nuclides above Fe, not reached by the $s$ and $r$ processes, are called $p$ nuclides. } $\iso{Mo}{92,94}$ and $\iso{Ru}{96,98}$, which are underproduced in other nucleosynthesis processes such as the $\gamma$ process in ccSN \citep{2011ApJ...729...46W,2013RPPh...76f6201R,2018ApJ...866..105B}.

Any conclusions on the importance of the \nup process depend not only on the choice of site but also on the amount of nuclides and the abundance pattern that can be produced in those sites. Therefore it is of great interest to study the uncertainties involved in the prediction of the resulting abundances, and especially which possible variation in the production is permitted by the uncertainties in the nuclear reaction rates used. On one hand, this allows the model uncertainties to be disentangled from the nuclear physics uncertainties, while on the other hand, it provides information on which isotope ratios are permitted because these depend on nuclear properties.

We have developed a Monte Carlo (MC) method allowing the variation of ten thousand rates simultaneously to address such questions \citep{2016MNRAS.463.4153R}. A simultaneous variation of rates is necessary to account for the combined action of rate changes. Neglection of such combinations may lead to an overemphasis of certain reactions and a misrepresentation of their impact on the total uncertainty \citep{2016MNRAS.463.4153R,2018AIPC.1947b0015R}. The method has been previously applied to investigate nucleosynthesis of $p$~nuclides in massive stars \citep{2016MNRAS.463.4153R} and in thermonuclear supernovae \citep{2018MNRAS.474.3133N}, and to study the weak $s$ process in massive stars \citep{2017MNRAS.469.1752N} and the main $s$ process in AGB stars \citep{2018MNRAS.478.4101C}. Here, we consistently extend our investigations to quantify the nuclear physics uncertainties in the synthesis of nuclides in the \nup process, applying a similar strategy and input as in the previous studies, and allowing a direct comparison of the resulting abundance uncertainties. Due to the fact that there is no single preferred site for the \nup process, a parameterisation of astrophysical conditions is used to cover a large range of possibilities.

The contents of the present paper are organised as follows. The parameterisation of the trajectories used in the MC approach is discussed in Section~\ref{sec:astrotracers}. The MC method itself is briefly presented in Section~\ref{sec:montecarlo}. The special importance of the \triplea and the $\iso{Ni}{56}({\rm n},{\rm p})\iso{Co}{56}$ reaction in the \nup process is discussed in Section~\ref{sec:triplea}. The results are shown and discussed in Section~\ref{sec:results} and a summary is given in Section~\ref{sec:summary}.

\section{Methods}
\label{sec:methods}

\subsection{Astrophysical models}
\label{sec:astrotracers}

\begin{figure}
\includegraphics[width=\columnwidth]{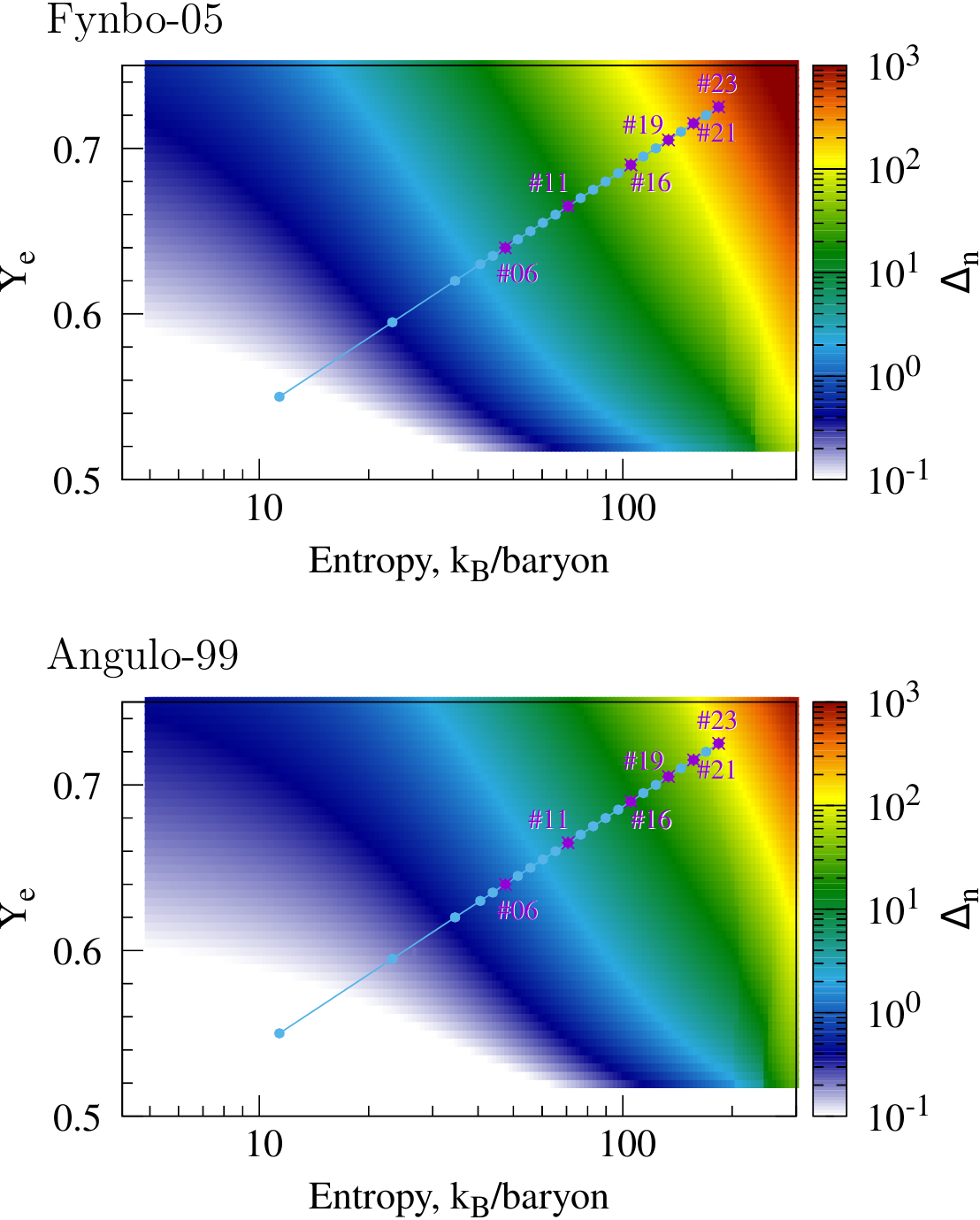}
\caption{\label{fig:modelspace}The explored parameter space in \ye and entropy $S$ for two choices of the \triplea rate. Dots correspond to trajectories used for the MC variations.}
\end{figure}

\begin{figure}
\includegraphics[width=\columnwidth]{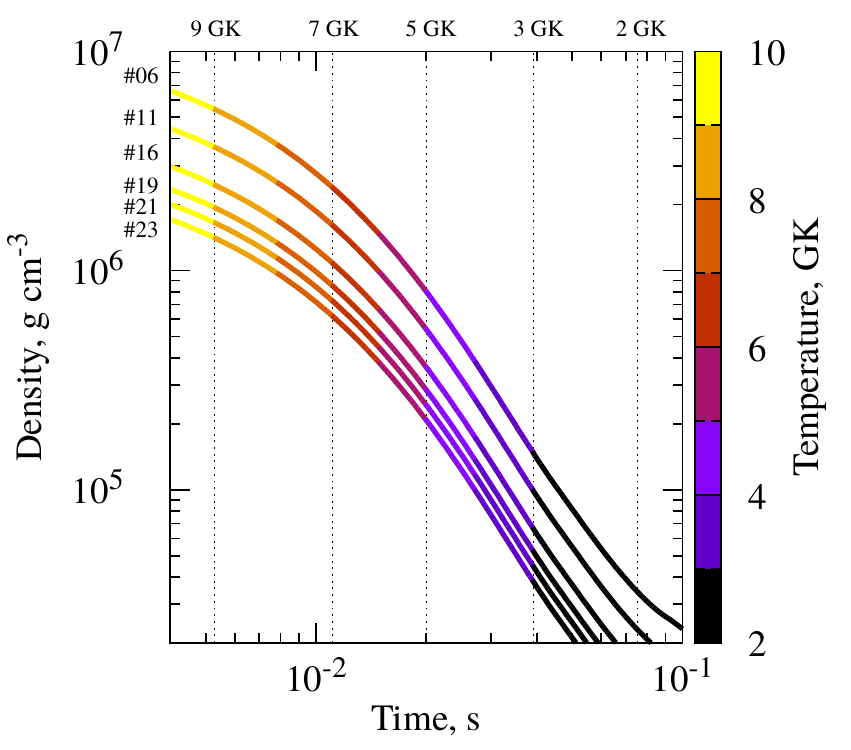}
\caption{\label{fig:trajectories}Time evolution (after the core bounce) of matter density for selected trajectories, based on the 
neutrino-driven wind component from PNS surface \citep{2012ApJ...758....9N}. The colour of each line shows the temperature at a 
given time.}
\end{figure}

\begin{table}
\centering
\caption{Initial conditions for each explored trajectory; the shown values of $\Delta_\mathrm{n}$ were obtained using the \triplea rate by \citet{2005Natur.433..136F} (Fynbo-05) and by \citet{1999NuPhA.656....3A} (Angulo-99), respectively. The six trajectories  labeled in Fig.~\ref{fig:modelspace} are underlined.\label{tab:conditions}}
\begin{tabular}{ccccc}
\hline
 Trajectory & \ye & Entropy & $\Delta_\mathrm{n}$ & $\Delta_\mathrm{n}$ \\
 &  & ($k_{\mathrm{B}}\;\rm{baryon}^{-1}$) & Fynbo-05 & Angulo-99\\
\hline
\#01 &  0.550 &   11.4 & $ 6.15\times 10^{-2}$ & $ 4.57\times 10^{-2}$ \\
\#02 &  0.595 &   23.2 & $0.356$ & $0.158$ \\
\#03 &  0.620 &   34.6 & $  1.15$ & $0.372$ \\
\#04 &  0.630 &   40.5 & $  1.89$ & $0.561$ \\
\#05 &  0.635 &   43.9 & $  2.43$ & $0.698$ \\
\underline{\#06} &  \underline{0.640} &   \underline{47.5} & \underline{3.13} & \underline{$0.873$} \\
\#07 &  0.645 &   51.5 & $  4.05$ & $  1.10$ \\
\#08 &  0.650 &   55.7 & $  5.22$ & $  1.40$ \\
\#09 &  0.655 &   60.3 & $  6.77$ & $  1.79$ \\
\#10 &  0.660 &   65.3 & $  8.74$ & $  2.30$ \\
\underline{\#11} &  \underline{0.665} &   \underline{70.7} & \underline{11.3} & \underline{$  2.97$} \\
\#12 &  0.670 &   76.6 & $  14.7$ & $  3.85$ \\
\#13 &  0.675 &   82.9 & $  19.0$ & $  4.99$ \\
\#14 &  0.680 &   89.7 & $  24.7$ & $  6.50$ \\
\#15 &  0.685 &   97.2 & $  32.0$ & $  8.50$ \\
\underline{\#16} &  \underline{0.690} &    \underline{105} & \underline{41.4} &  \underline{$11.1$} \\
\#17 &  0.695 &    114 & $  53.7$ & $  14.6$ \\
\#18 &  0.700 &    123 & $  69.4$ & $  19.1$ \\
\underline{\#19} &  \underline{0.705} &    \underline{134} & $\underline{89.6}$ & \underline{$24.9$} \\
\#20 &  0.710 &    145 & $ 1.17\times 10^{2}$ & $  32.6$ \\
\underline{\#21} &  \underline{0.715} &    \underline{157} & \underline{$1.63\times 10^{2}$} & \underline{$42.6$} \\
\#22 &  0.720 &    169 & $ 2.23\times 10^{2}$ & $  58.0$ \\
\underline{\#23} &  \underline{0.725} &    \underline{184} & \underline{$3.05\times 10^{2}$} &\underline{ $84.7$} \\
\hline
\end{tabular}
\end{table}

The efficiency of \nup-process nucleosynthesis depends on the detailed conditions encountered in the neutrino wind. Among the crucial parameters are initial composition, matter density, and temperature of the ejecta, as well as their expansion rate (determining the time evolution of matter density and temperature) and neutrino-wind properties. Since these conditions, on one hand, are not constrained well by current ccSN explosion models \citep{2018ApJ...855..135B} and, on the other hand, a range of conditions is expected to occur either within one site or in different sites, we investigated a large range of possible environments.

Similar to the ratio of neutron abundance to seed abundance in the $r$ process, the number ratio $\Delta_{\rm n}$ of free neutrons, created by the reaction ${\rm p}(\overline{\nu}_\mathrm{e}, {\rm e}^+){\rm n}$, and seed nuclei  is a good indicator for the strength of the \nup process, as introduced by \citet{pruet06}. It is given by
\begin{equation}
\label{eq:deltan}
\Delta_{\rm n} \equiv \frac{Y_{\rm p}}{Y_{\rm h}} n_{\bar{\nu}_{\rm e}}
= \frac{Y_{\rm p}}{Y_{\rm h}} \int_{T_9 \leq 3} \lambda_{\bar{\nu}_{\rm e}} {\rm d}t\ ,
\end{equation}
where $\lambda_{\bar{\nu}_{\rm e}}$ is the rate for ${\rm p} + \nu_e \rightarrow {\rm n} + {}^+{\rm e}$ and $Y_\mathrm{h}$ is the seed abundance, i.e., the abundance of nuclei with $Z>2$, taken at the onset of the \nup process at $T_9=3$. The seed abundance is in large part determined by the abundance of \iso{Ni}{56}. A detailed discussion of the significance of $\Delta_{\rm n}$ is found in \citet{2011ApJ...729...46W}.

We used a set of parameterised models covering electron fractions of $0.55 \leq \ye \leq 0.725$ and entropies of $11.4\leq S\leq 184$ $k_{B}$~baryon$^{-1}$, taken as initial values at the time of freeze-out from NSE at 7 GK. The choice of $\ye$ and entropy also determines $\Delta_{\rm n}$. As illustrated in Fig.~\ref{fig:modelspace}, within these ranges we probe an extensive set of $\Delta_{\rm n}$ values allowing for a \nup process, from the most feeble onset to strong processing of heavier nuclei. The evolution of temperature and density is based on a typical PNS wind trajectory from a 1D neutrino-hydrodynamics simulation \citep[see,][and references therein]{2012ApJ...758....9N}. Adopting the temperature evolution of the original trajectory, we adjusted the density by multiplying it with a factor consistent with a given entropy.

Examples of the obtained density and temperature as function of time for a few selected trajectories are shown in Fig.~\ref{fig:trajectories}. In the nucleosynthesis calculations, we only took into account neutrino absorption on nucleons, which is mainly $\bar{\nu_{\rm e}} + {\rm p} \rightarrow {\rm n} + {\rm e}^{+}$. The neutrino properties are consistent with the hydrodynamical evolution of a PNS: The values of the luminosity and the mean energy for the anti-electron neutrino are $L_{\bar{\nu_{\rm e}}} = 2.06 \times 10^{51}~{\rm erg}$ and $\epsilon_{\bar{\nu_{\rm e}}}=15.2~{\rm MeV}$, respectively, at the beginning of the nucleosynthesis calculations (at 7 GK). The $\ye$ did not change significantly (only decreased by $\sim 0.005$) between the end of NSE and the end of the \nup nucleosynthesis. The details of the trajectories used in the MC study are also summarised in Table~\ref{tab:conditions}.

\subsection{Nucleosynthesis with Monte Carlo variations}
\label{sec:montecarlo}

\begin{figure}
\includegraphics[width=\columnwidth]{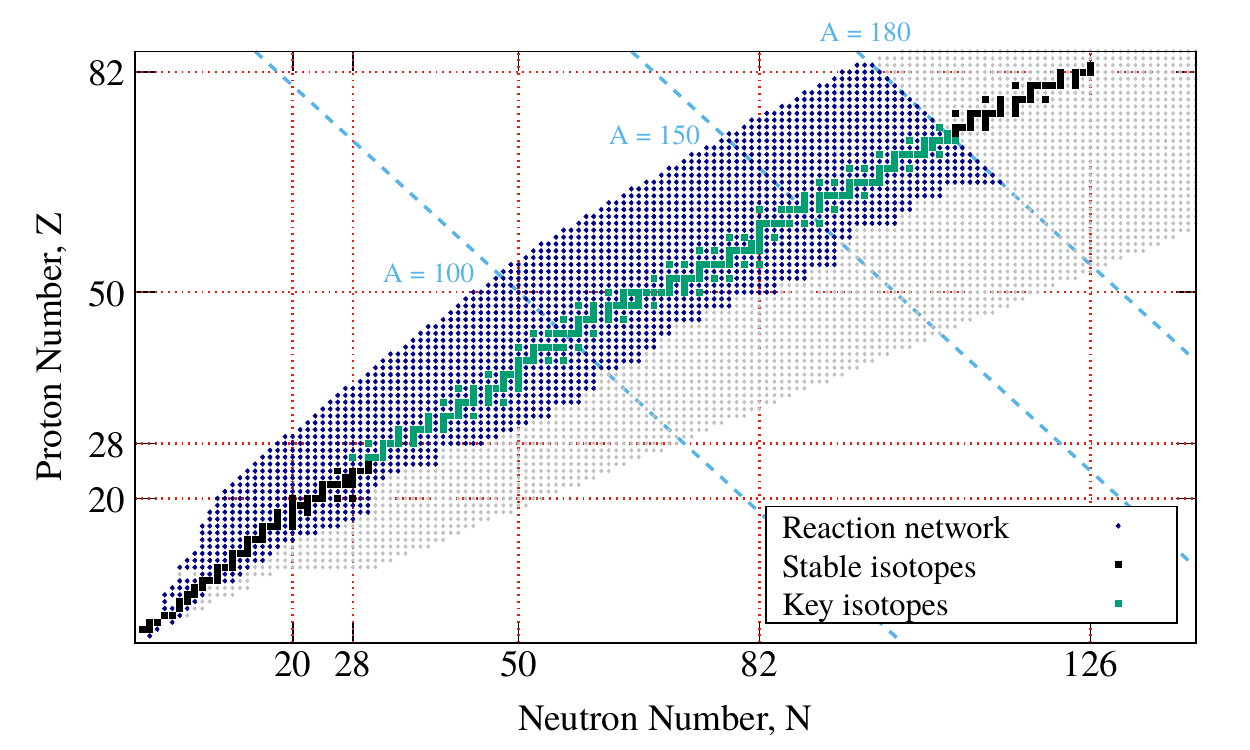}
\caption{\label{fig:np_network}Nuclides included in the reaction network on the $N$--$Z$ plane.}
\end{figure}

The trajectories (see Section~\ref{sec:astrotracers}) were post-processed using the \textsc{PizBuin} code suite, consisting of a fast reaction network and a parallelised Monte Carlo driver. Our reaction network calculations started at $T= 7$ GK and followed the nucleosynthesis throughout the freeze-out and final decay back to stability. We used the same procedure as presented in detail in \citet{2016MNRAS.463.4153R} and previously applied to various further nucleosynthesis sites (see Section~\ref{sec:intro}). Therefore only the main points of the procedure are very briefly summarized here.

The reaction network contained 2,216 nuclides, including nuclides around stability and towards the proton-rich side of the nuclear chart, as shown in Fig.~\ref{fig:np_network}. The standard rate set and the assigned uncertainties were the same as previously used in our works \citep{2016MNRAS.463.4153R, 2017MNRAS.469.1752N, 2018MNRAS.474.3133N, 2018MNRAS.478.4101C}: rates for neutron-, proton-, and $\alpha$-induced reactions were a combination of theoretical values by \citet{2000ADNDT..75....1R}, supplemented by experimental rates taken from \citet{2006AIPC..819..123D} and \citet{2010ApJS..189..240C}. Decays and electron captures were taken from a REACLIB file compiled by \citet{freiburghaus1999} and supplemented by rates from \citet{1987ADNDT..36..375T} and \citet{1999A&A...342..881G} as provided by \citet{2005A&A...441.1195A} and \citet{2013NuPhA.918...61X}.

Each trajectory was run 10,000 times in a network calculation, with each rate subject to a different rate variation factor for each run. The combined output was then analyzed. For each trajectory, the total uncertainty in the final abundances after decay to stability was calculated and key rates, i.e. those that dominate the uncertainty of a given final isotopic abundance, were identified. By our definition, reducing the uncertainty of a key rate will also considerably decrease the uncertainty in a final abundance. The identification of key rates was achieved by examining the correlation between a change in a rate and the change of an abundance, as found in the stored Monte Carlo data. As before, the Pearson product-moment correlation coefficient \citep{doi:10.1098/rspl.1895.0041} was used to quantify correlations. The Pearson correlation coefficient $r_\mathrm{cor}$ can assume values $ 0 \leq \left|r\right| \leq 1$. Positive values of $r_\mathrm{cor}$ indicate a direct correlation between rate change and abundance change, whereas negative values signify an inverse correlation, i.e., the abundance decreases when the rate is increased. The larger the absolute value of the Pearson coefficient, the stronger the correlation. As in our previous work, a key rate was identified by $\left|r_\mathrm{cor}\right|\geq 0.65$.

Each astrophysical reaction rate on target nuclides from Fe to Bi was varied within its own uncertainty range. Forward and reverse rates received the same variation factor as they are connected by detailed balance. The assigned uncertainty range is temperature dependent and constructed from a combination of the measured uncertainty (if available) for target nuclei in their ground states and a theory uncertainty for predicted rates on nuclei in thermally excited states. Theory uncertainties were different depending on the reaction type and can be asymmetric. Details are given in \citet{2016MNRAS.463.4153R,2018AIPC.1947b0015R}. In the present context it is important to note that the nucleosynthesis path is located a few units away from stability and therefore there are no experimentally determined reaction rates available (except for the \triplea and a few reactions acting on stable nuclides at late times, see Section~\ref{sec:results}). Furthermore, the temperatures in the \nup process are so high that reactions on thermally excited states of nuclei dominate the reaction rate \citep{2012ApJS..201...26R, rauadvance} and these are not constrained experimentally. Thus, effectively the uncertainties in the reaction rates were dominated by the assumed theory uncertainties. For example, the two most important reaction types, (n,p)$\leftrightarrow$(p,n) and (p,$\gamma$)$\leftrightarrow$($\gamma$,p), were varied from $1/3$ the standard rate to twice the standard rate and (p,$\alpha$)$\leftrightarrow$($\alpha$,p) rates were varied between $1/10$ and twice the standard rate.

The present MC study does not include uncertainties on nuclear masses. Nevertheless, it is worth noting that uncertainties in the nuclear masses affect the equilibrium abundances within an isotonic chain established by the (p,\,$\gamma$)$-$($\gamma$,\,p) equilibrium (see Secs.\ \ref{sec:intro} and \ref{sec:generalfeatures}) because they change the ratio of forward and reverse reaction. Compared to the situation in the $rp$ process, however, uncertainties in mass differences, which affect the proton separation energies, are of lesser importance in the $\nup$ process. This is due to the different hydrodynamical conditions, the dominance of fast (n,\,p) reactions over competing proton captures or $\beta^+$ decays, and the different location of the $\nup$-process path, proceeding closer to stability and involving fewer nuclides with inaccurately determined masses. \citet{2011ApJ...729...46W} quotes a number of nuclides for which nuclear masses should be determined with smaller uncertainty. A number of experimental investigations have targeted masses of nuclides in the \nup-process path \citep[see, e.g.,][]{PhysRevC.78.054310,PhysRevLett.106.122501,XING2018358}.

\section{The features of \nup-process nucleosynthesis}
\label{sec:features}

\subsection{General}
\label{sec:generalfeatures}

A \nup process can occur in proton-rich, hot ejecta expanding in a flow of anti-electron neutrinos ($\overline{\nu}_\mathrm{e}$). The ejecta quickly cool from the initially very high temperature, at which time only nucleons were present. In the first phase of the cooling nucleons are assembled mainly to $\iso{Ni}{56}$ and $\alpha$-particles in a nuclear statistical equilibrium, leaving a large number of free protons. At sufficiently low temperature ($\leq 3-4$ GK), rapid proton captures ensue on $\iso{Ni}{56}$. Production of heavier nuclei would be stopped at $\iso{Ge}{64}$, which has an electron-capture lifetime longer than a minute. This is too long in comparison with the expansion timescale (of the order of seconds) to allow for production of an appreciable number of nuclides beyond $\iso{Ge}{64}$ before nuclear reactions freeze out. In the \nup process, however, a small number of free neutrons are continuously created by $\overline{\nu}_\mathrm{e}$ captures on the free protons. This supply of free neutrons allows for (n,\,p) reactions bypassing any slow electron captures and $\beta^+$ decays, not just of $\iso{Ge}{64}$, but also of
other potential bottlenecks at higher mass number.

The main nucleosynthesis flow in the \nup process is characterised by rapid proton captures in a (p,\,$\gamma$)-($\gamma$,\,p) equilibrium with (n,\,p) reactions connecting the contiguous isotonic chains. Although such an equilibrium is also achieved in the $rp$ process on the surface of accreting neutron stars \citep{schatz}, the \nup process proceeds at lower density than the $rp$ process. The resulting nucleosynthesis path follows the $N = Z$ line only up to the Mo region, reaching further and further into neutron-richer isotopes between Mo and Sn, moving gradually away from the $N = Z$ line \citep{2011ApJ...729...46W}. The path is pushed strongly towards stability at the Sn isotopes and above, providing a strong barrier for the efficient production of any elements beyond Sn. Decay and (n,\,p) reaction timescales are longer for nuclides closer to stability
and the higher Coulomb barriers suppress proton captures.

The location of the effective $\nu$p-process path is determined by the nuclear properties giving rise to the (p,\,$\gamma$)-($\gamma$,\,p) equilibrium and the very fast (n,\,p) reactions, and remains remarkably unaffected by variations of the astrophysical parameters within realistic limits such as entropy, $\ye$, and expansion timescale, as long as the conditions permit the appearance of a \nup process. Whenever a \nup process occurs, the nucleosynthesis path beyond \iso{Ni}{56} initially follows the $N=Z$ line and gradually veers off towards stability. Systematic variations of reaction rates show only small effects, if any, regarding the path location. This is a consequence of the (p,\,$\gamma$)-($\gamma$,\,p) equilibrium in which the path is determined by nuclear mass differences \citep{schatz}. All these variations, however, determine how far up the path follows the $N=Z$ line before diverging, or whether it is terminated already at low charge numbers, $Z$. Consequently, it is clear that the achieved abundances within the path are also determined by these conditions. This motivates the introduction of the quantity $\Delta_\mathrm{n}$ as defined in Eq.\ (\ref{eq:deltan}).

On the nuclear reaction side, it is expected that the results are mostly insensitive to proton captures due to the prevailing (p,\,$\gamma$)-($\gamma$,\,p) equilibrium. Only at late freeze-out times does this equilibrium break down, giving rise to some sensitivity to a variation of rates. There may also be some sensitivity to proton captures located at the end of the nucleosynthesis path that are not, or only barely, in equilibrium.

\begin{figure}
\includegraphics[width=\columnwidth]{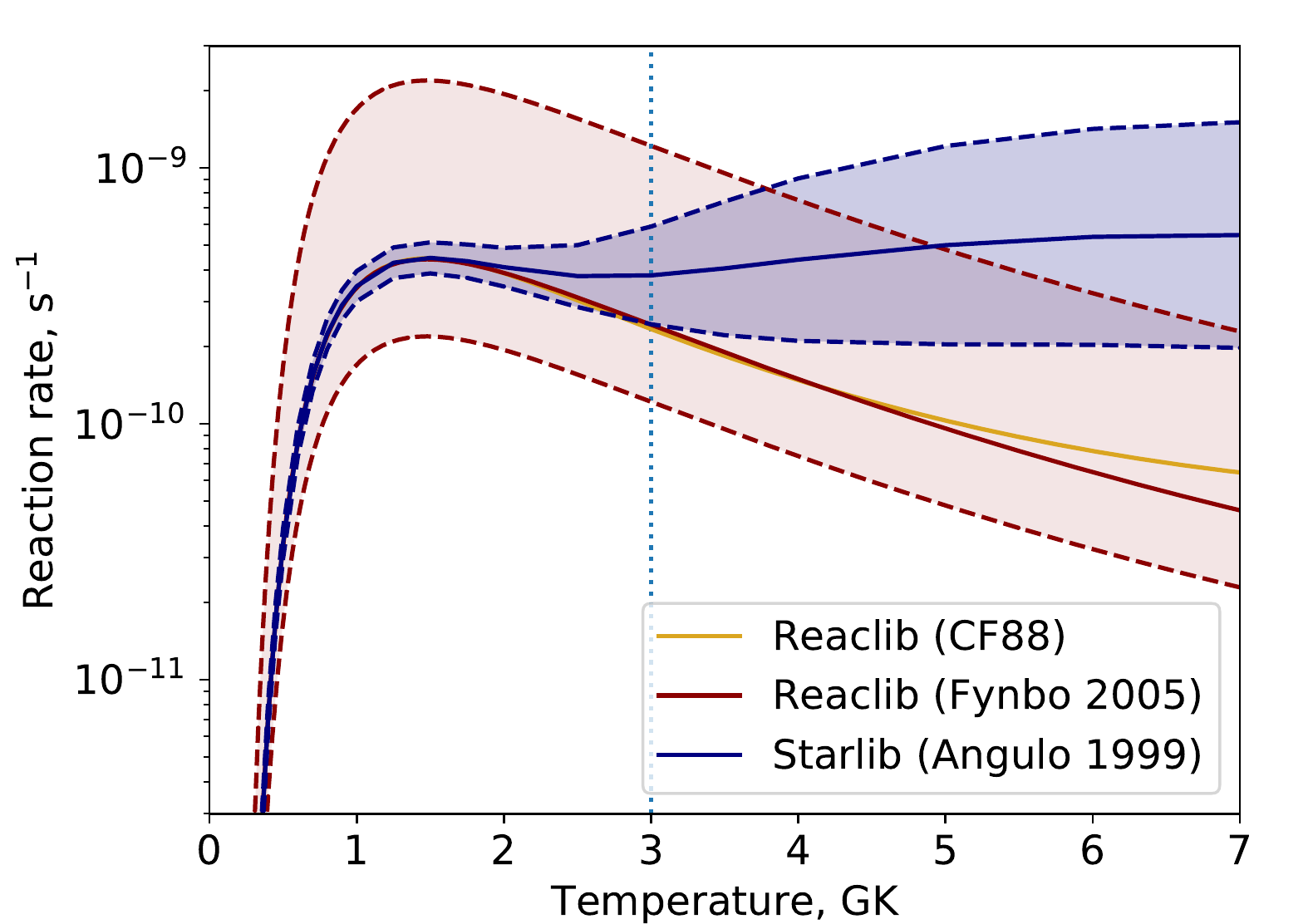}
\caption{\label{fig:triplea}Comparison of \triplea rates as a function of temperature. The uncertainty factor assigned to the rate of \citet{2005Natur.433..136F} was $\times$5 upwards and $\times$0.5 downwards. The reaction rate by \citet{1999NuPhA.656....3A} was adopted with the {\tt Starlib} uncertainty evaluation. The older standard rate by \citet{1988ADNDT..40..283C} (CF88) is also plotted. It is close to the \citet{2005Natur.433..136F} rate at low temperature.}
\end{figure}

\begin{figure*}
\includegraphics[width=2\columnwidth]{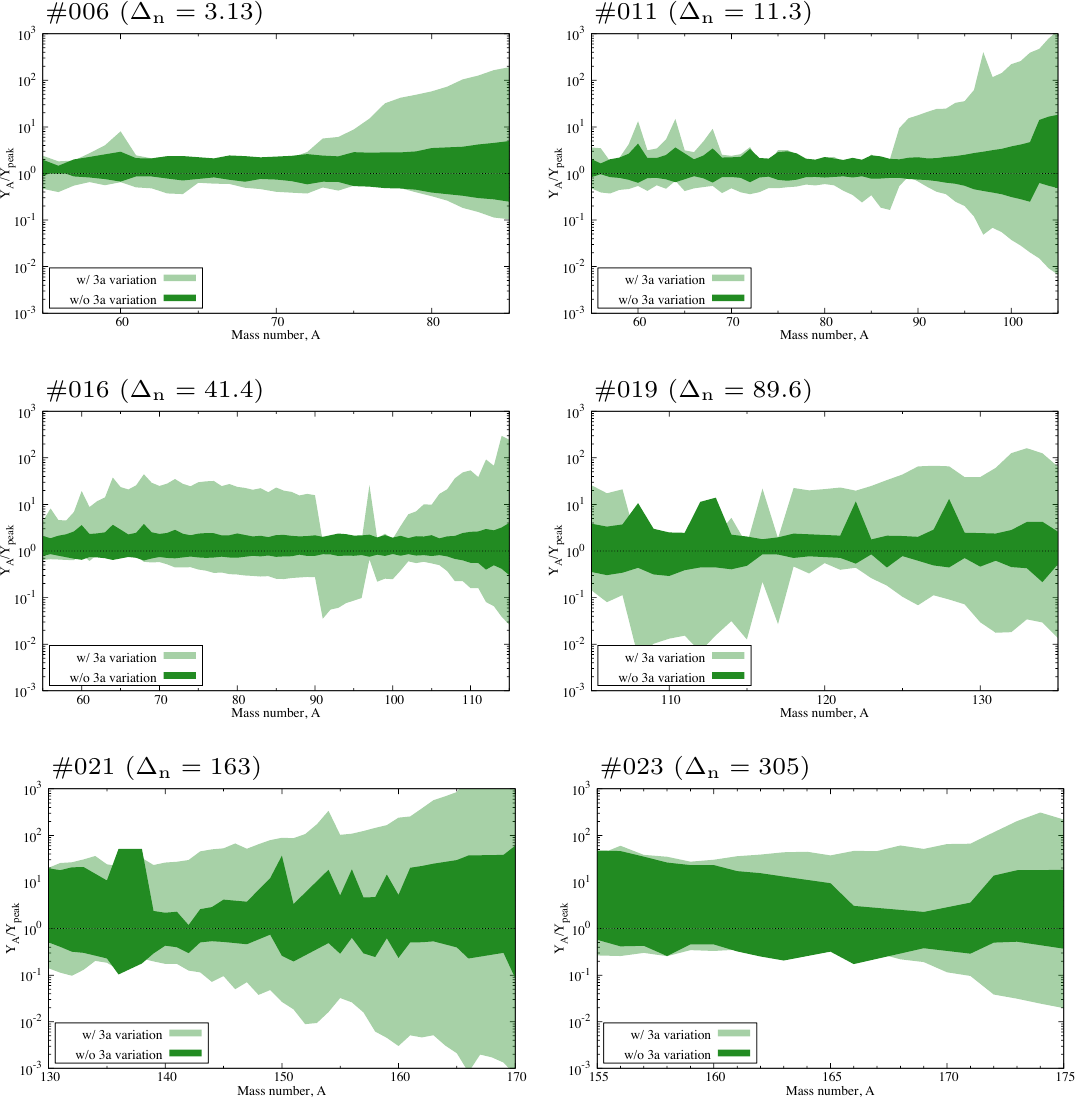}
\caption{Final uncertainties obtained in six selected trajectories with and without variation of the \triplea rate. The colour shade corresponds to a 90\% probability interval for each isobaric abundance ($Y_{A}$), normalized to the peak value ($Y_{\rm peak}$).\label{fig:triplevariation}}
\end{figure*}

The flow to heavier nuclei is determined by (n,\,p) reactions and thus a knowledge of these is essential. For a given choice of astrophysical conditions, faster (n,\,p) rates result in processing further up the nucleosynthesis path. Whether a given (n,\,p) reaction is important, however, depends on whether its target nucleus is actually in the path and whether it receives an appreciable abundance as given by the (p,\,$\gamma$)-($\gamma$,\,p) equilibrium. Neutron captures on proton-rich nuclei may be of some relevance at large $Z$ and/or at late times, depending on the hydrodynamic evolution of the trajectory \citep{2011ApJ...729...46W,arcofromart}.

A special class of reactions are those which govern the onset of the \nup process at high temperature. When freezing out from nuclear statistical equilibrium at high temperature, the \nup process is delayed by several factors. At high temperature, ($\gamma$,\,p) reactions are fast and the equilibrium abundances are always located around $^{56}$Ni. Since the main abundance is concentrated in $^{56}$Ni, further processing is halted until the $^{56}$Ni waiting point can be bridged effectively and the (p,\,$\gamma$)-($\gamma$,\,p) equilibrium abundance maxima in the subsequent isotonic chains are moved to higher $Z$. This depends on the competing rates of ($\gamma$,\,p), (n,\,$\gamma$), and (n,\,p) on $^{56}$Ni and occurs at $T\approx 3.5$ GK.

Whether further processing occurs at this temperature depends on the relative speeds of $(\gamma, \alpha)$, $({\rm p}, \alpha)$, and $({\rm n}, \alpha)$ reactions on waiting point isotopes of Zn and Ge compared to the $({\rm n}, {\rm p})$, $({\rm n}, \gamma)$, or $({\rm p}, \gamma)$ reactions required to commence the nucleosynthesis to heavier elements. It has been shown that reaction cycles can form via $({\rm n}, \alpha)$ or $({\rm p}, \alpha)$ reactions and further delay the processing to heavier mass \citep{arcofromart,rauadvance}. Since these depend on competitions between particle-induced reactions, they do not depend strongly on the time-dependence of the density imposed by a chosen trajectory. A modification of the density at a given temperature affects proton- and neutron-induced reactions similarly and only changes the relation between proton captures and $(\gamma, {\rm p})$ reactions. The strongest dependence on an astrophysical parameter is the one on $Y_\mathrm{n}$ created by the $\overline{\nu}_\mathrm{e}$ flux present at a given temperature. However, this does not change the ratio between $({\rm n}, \gamma)$, $({\rm n}, {\rm p})$, and $({\rm n}, \alpha)$ reactions, the latter being a hindrance to the flow up to heavier nuclei. Another important aspect is the time evolution of the trajectory because it determines for how long favorable conditions for a cycle (if existing) are upheld.

In our MC variation study, we do not explicitly inspect reaction flows but, of course, the above cases are accounted for in the network runs automatically and thus are implicitly included in the analysis of final abundances and key reactions given in Section~\ref{sec:results}.

\subsection{Importance of the ``bottleneck'' reactions: 3$\alpha$ and $\iso{Ni}{56}$(n,\,p)$\iso{Co}{56}$}
\label{sec:triplea}

While the Monte Carlo variations focus on reactions on Fe isotopes and above, it is important to note that the efficiency of \nup-process nucleosynthesis strongly depends on the \triplea (the two-step reaction with the first step being $\iso{He}{4} + \iso{He}{4} \rightarrow \iso{Be}{8}$ immediately followed by $\iso{Be}{8} + \iso{He}{4} \rightarrow \iso{C}{12}$), which thus is an important key reaction. It is never in equilibrium and determines the relative abundance of $\alpha$ particles, protons, and \iso{Ni}{56} at the onset and during the \nup process. It therefore determines the $\iso{Ni}{56}$ seed available for further processing up to heavier masses and thus also plays a dominant role in the production of heavy nuclei. Despite of the importance of this reaction, the \triplea bears a large experimental uncertainty in the high temperature regime as well as in the lower temperature region, the latter being mainly important for stellar evolution.

\begin{figure}
\includegraphics[width=\columnwidth]{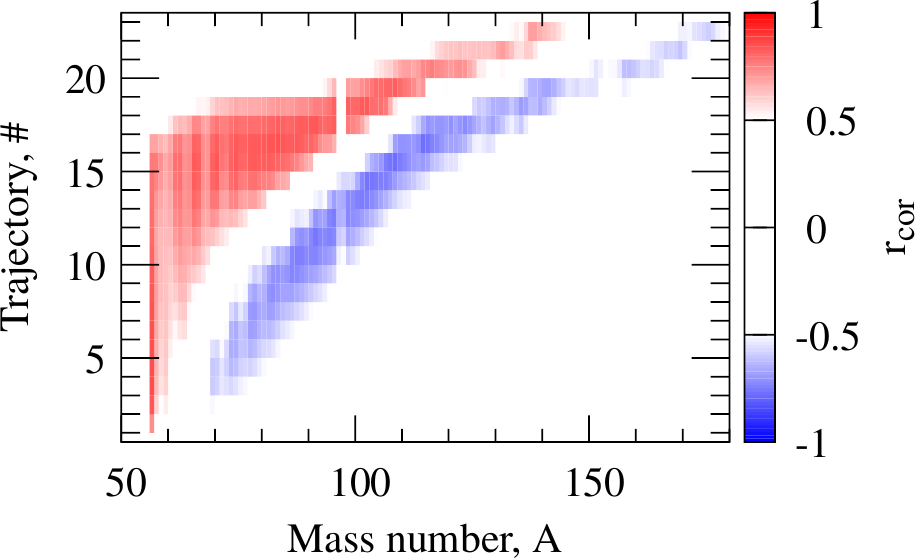}
\caption{The impact of the {\triplea} rate \citep{2005Natur.433..136F} on the production of nuclides for all trajectories. Shown is the correlation of the abundance variation of a given nuclide with the variation of the \triplea rate.\label{fig:tripleall}}
\end{figure}

Fig.~\ref{fig:triplea} presents the \triplea rates, together with their uncertainties, as determined by \citet{2005Natur.433..136F} (as given in the JINA REACLIB) and \citet{1999NuPhA.656....3A} \citep[as given in][]{2013ApJS..207...18S}. The older rate of \citet{1988ADNDT..40..283C} (also given in the JINA REACLIB) is also shown. In Fig.~\ref{fig:triplevariation} we show the final MC-computed abundances, and their uncertainties, obtained with the \triplea rate of \citet{2005Natur.433..136F} and its uncertainty as given in Fig.~\ref{fig:triplevariation}, for the trajectories \#06, \#11, \#16, \#19, \#21, and \#23 (see Table~\ref{tab:conditions}). The impact of the \triplea rate on the production of nuclides in all trajectories is summarised in Fig.~\ref{fig:tripleall}. As becomes obvious in Fig.~\ref{fig:triplevariation}, the variation in final abundances is so strong that it would cover most variations caused by uncertainties in rates involving nuclides heavier than Fe. Therefore we chose a \enquote{standard} rate for the \triplea and did not vary it further during the MC procedure. Our ``standard'' rate is the one of \citet{2005Natur.433..136F} as given in the JINA REACLIB. 

Fig.~\ref{fig:modelspace} and Table~\ref{tab:conditions} provide $\Delta_{\rm n}$ values for the two choices of \triplea rates. As can be seen easily in Table~\ref{tab:conditions} the choice of \triplea rate affects at which initial conditions a specific value of $\Delta_\mathrm{n}$ is achieved. For example, using the \citet{2005Natur.433..136F} rate a value of $\Delta_\mathrm{n}\approx 19$ is found in trajectory \#13 whereas a similar value is found in trajectory \#18 for the \citet{1999NuPhA.656....3A} rate. This explains why the overall production patterns are shifted in Fig.~\ref{fig:finalabuns} when comparing the results obtained with these two rates. Trajectories with larger $\Delta_{\rm n}$ produce heavier nuclei because with a larger supply of neutrons the nucleosynthesis path can run further up to larger mass numbers. A slower \triplea rate leaves more protons at the onset of the processing and thus reduces the \iso{Ni}{56} seed.

\citet{2011ApJ...729...46W} identified two reaction sequences competing with the \triplea. These sequences are determined by the reactions $\iso{Be}{7}$($\alpha$,\,$\gamma$)$\iso{C}{11}$ and $\iso{B}{10}$($\alpha$,\,p)$\iso{C}{13}$. Their uncertainties have a similar impact as the one in the \triplea discussed above.

\begin{figure}
\includegraphics[width=\columnwidth]{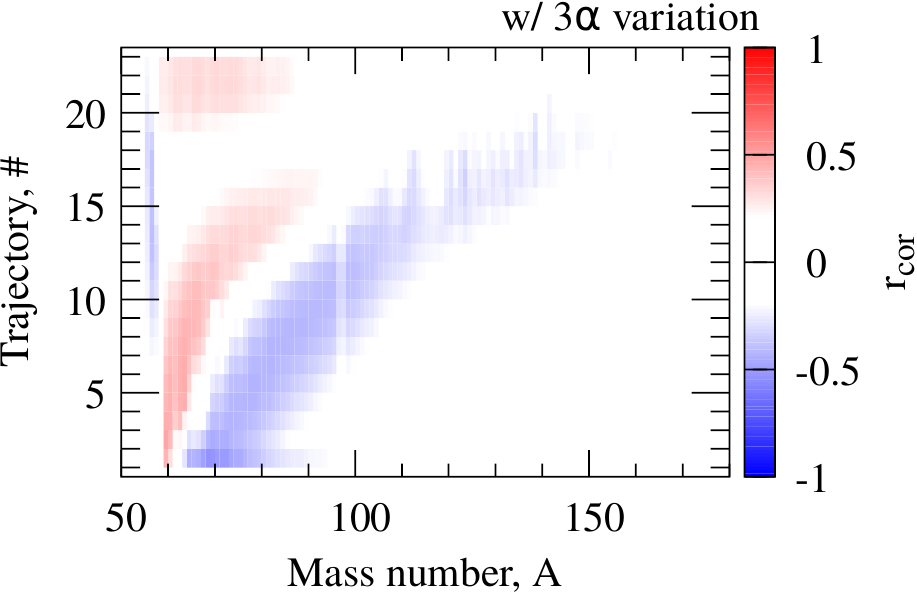}
\vspace{5pt}

\includegraphics[width=\columnwidth]{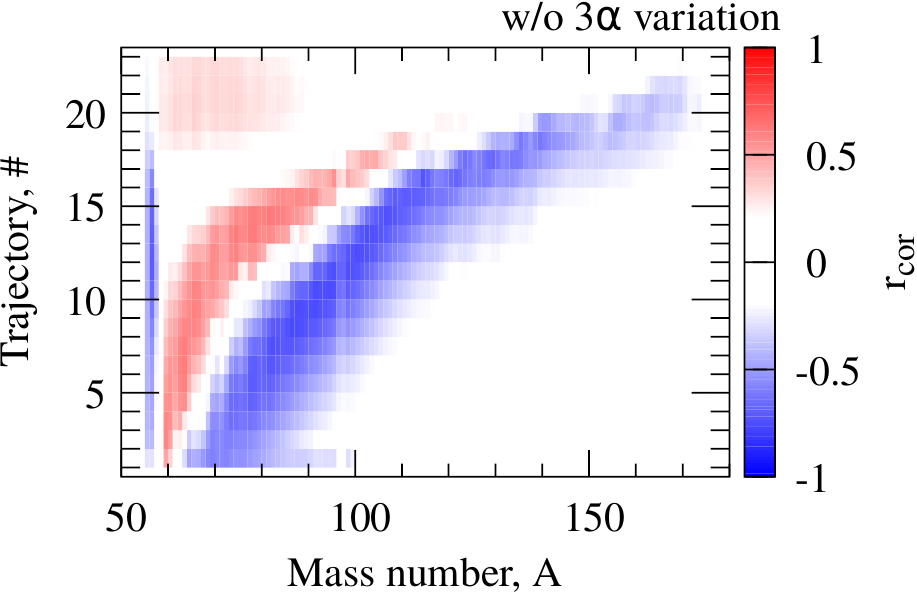}
\caption{The impact of the $\iso{Ni}{56}({\rm n},{\rm p})\iso{Co}{56}$ rate on the production of nuclides for all trajectories. Shown is the correlation of the abundance variation of a given nuclide with the variation of the $\iso{Ni}{56}({\rm n},{\rm p})\iso{Co}{56}$ reaction rate, with (top panel) and without (bottom panel) simultaneous variation of the \triplea rate.\label{fig:ni56np}}
\end{figure}

\begin{figure}
\includegraphics[width=\columnwidth]{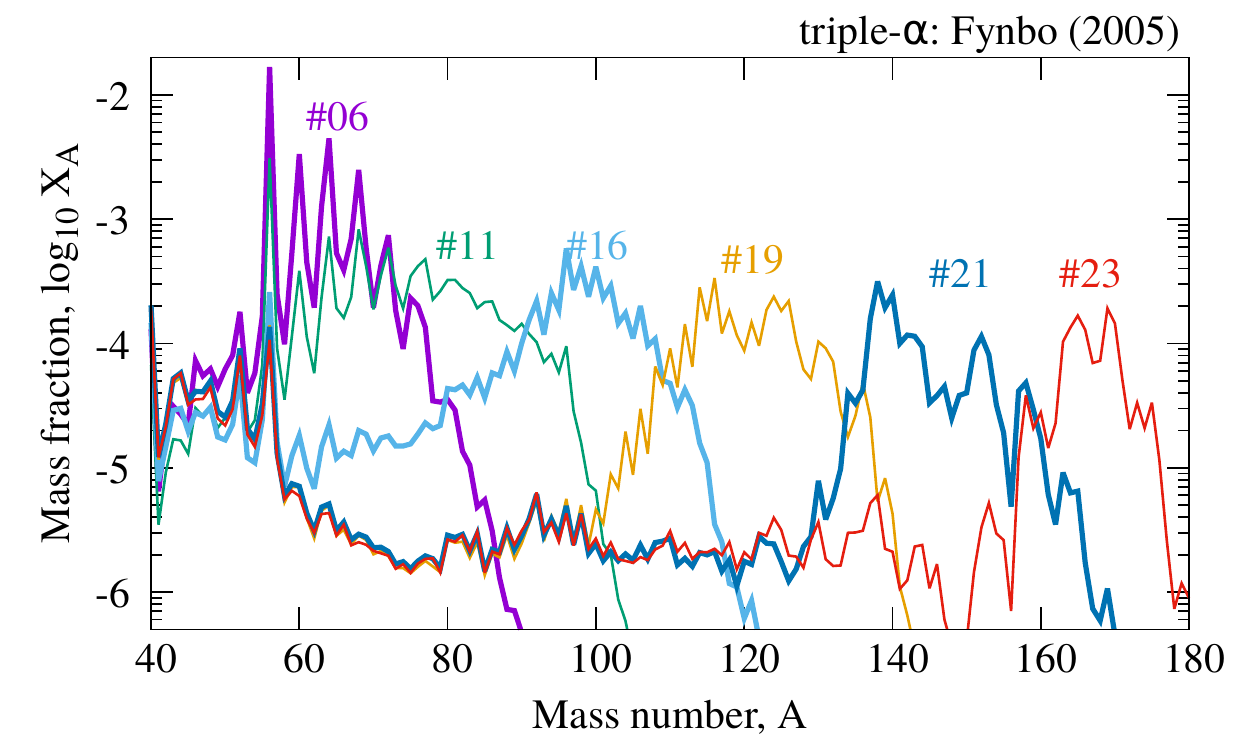}
\includegraphics[width=\columnwidth]{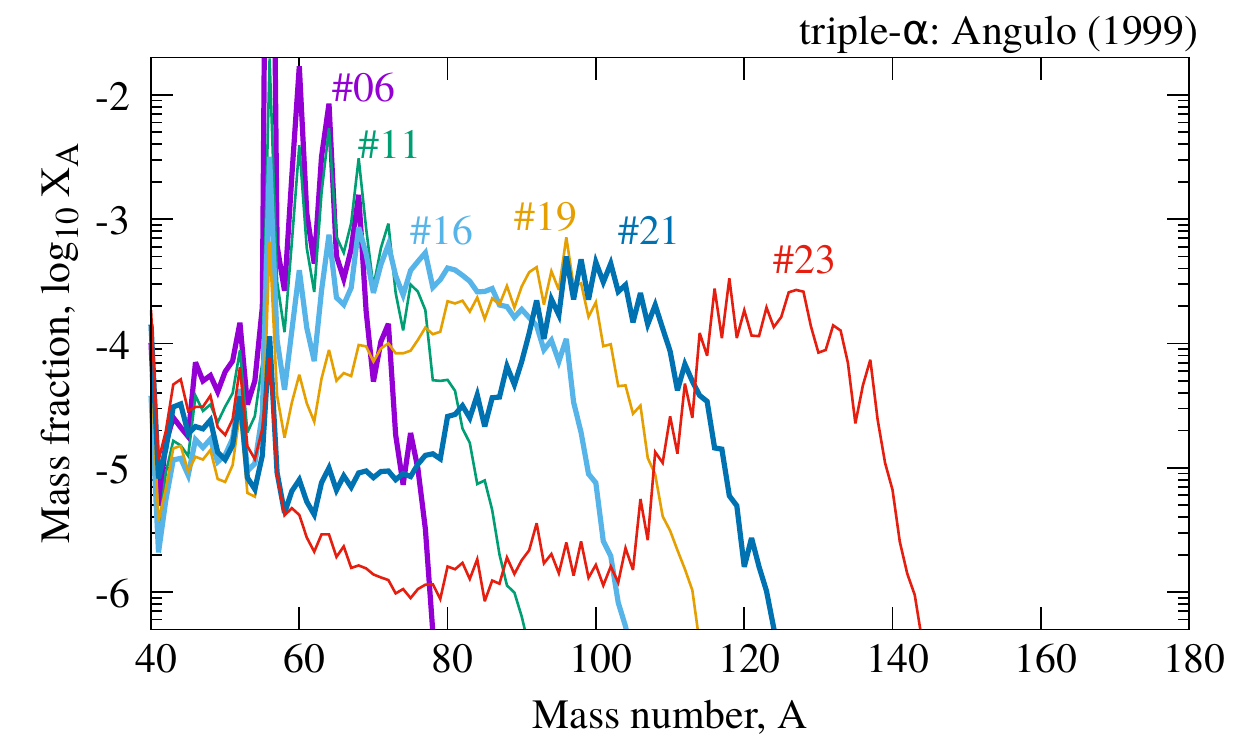}
\caption{\label{fig:finalabuns}Final mass fractions obtained in selected trajectories and with two different \triplea rates. All other rates have not been varied but kept at their standard values.}
\end{figure}

\begin{figure*}
\includegraphics[width=0.95\hsize]{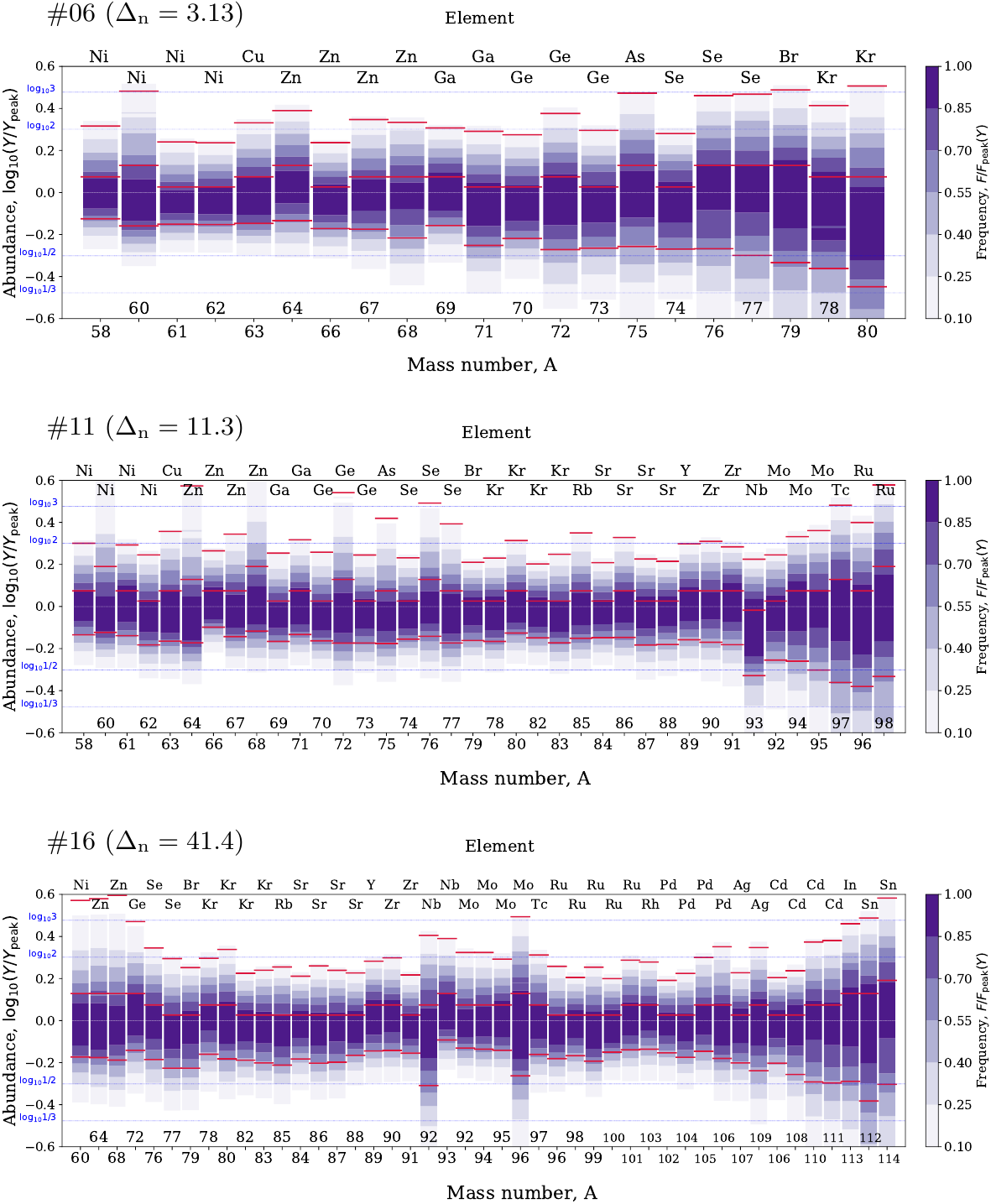}
\caption{Total production uncertainties of stable nuclei due to rate uncertainties in the MC post-processing of the trajectories \#06, \#11, and \#16. The colour shade gives the relative probabilistic frequency $Y/Y_{\rm peak}$ (final abundances $Y$ normalized to the peak value $Y_{\rm peak}$) and the horizontal red lines mark cumulative frequencies of 5\%, 50\%, and 95\% for each distribution. Uncertainty factors of two and three are marked by horizontal lines in blue. Note that the uncertainties are asymmetric and that the abundance scale is logarithmic.\label{fig:uncertall1}}
\end{figure*}

\begin{figure*}
\includegraphics[width=0.95\hsize]{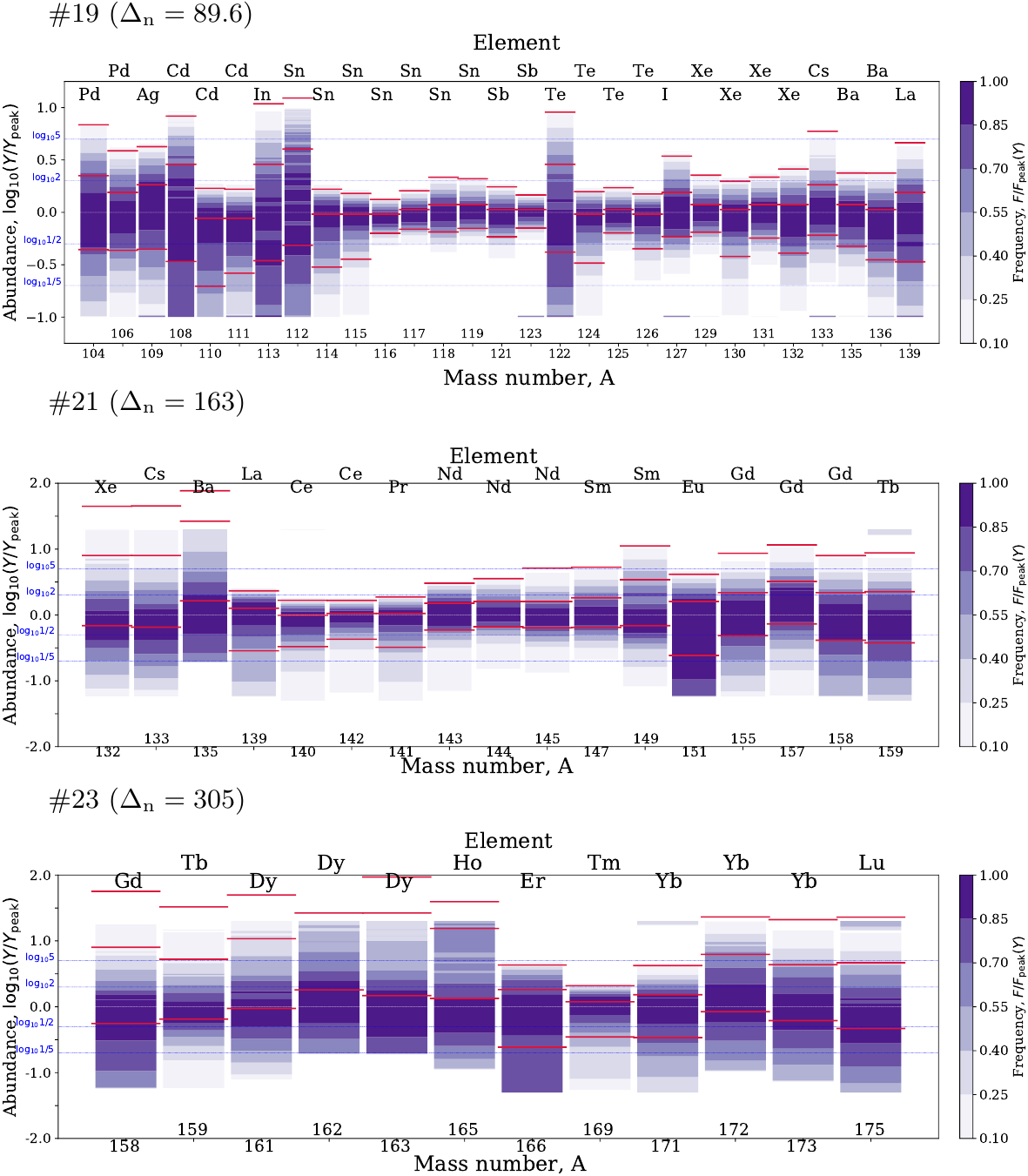}
\caption{Same as Fig.~\ref{fig:uncertall1} but for trajectories \#19, \#21, and \#23.\label{fig:uncertall2}}
\end{figure*}

\begin{figure*}
\includegraphics[width=2\columnwidth]{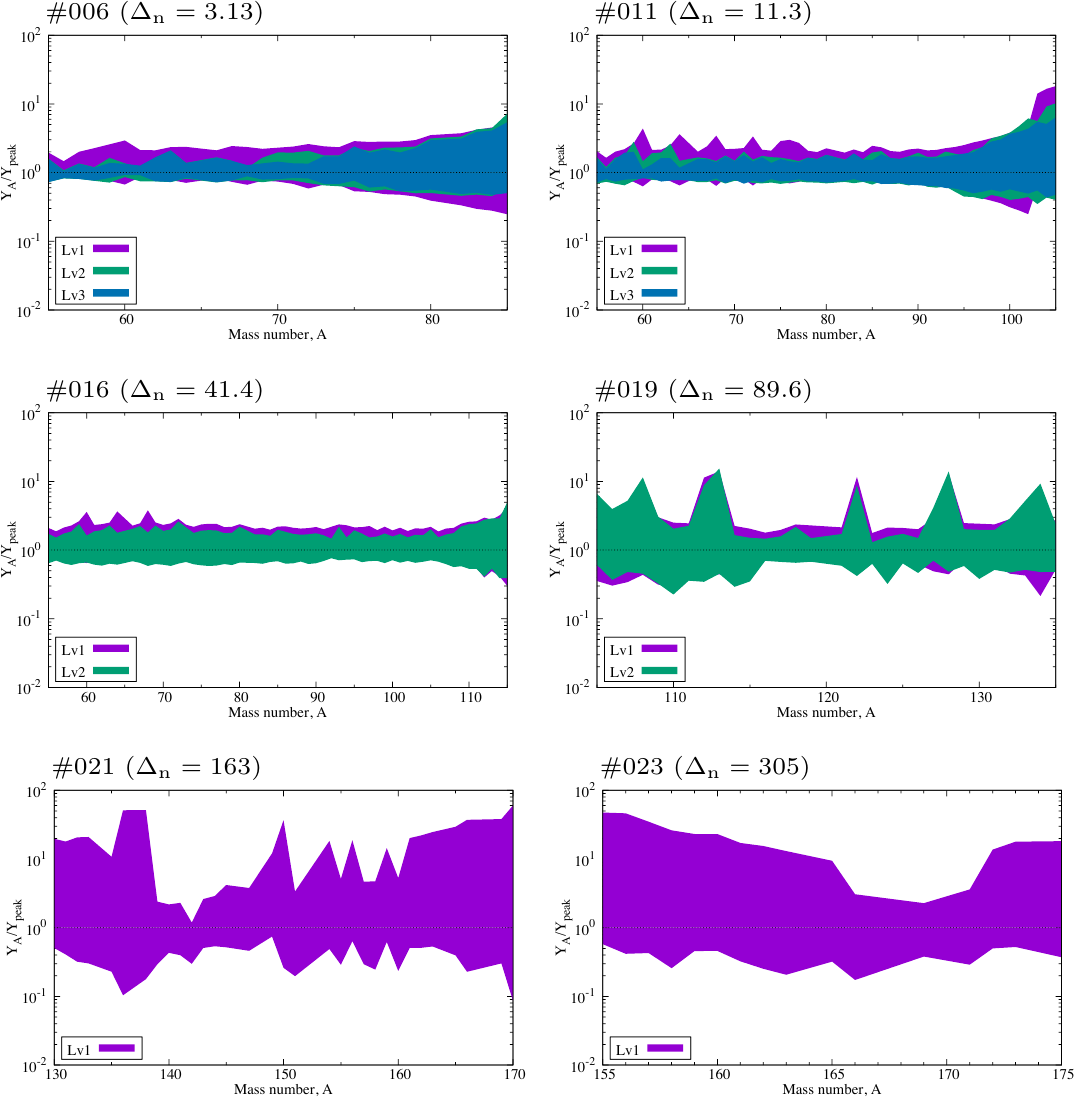}
\caption{\label{fig:keylevels}Final uncertainties obtained in six selected trajectories for several levels.}
\end{figure*}

\begin{figure*}
\includegraphics[width=2\columnwidth]{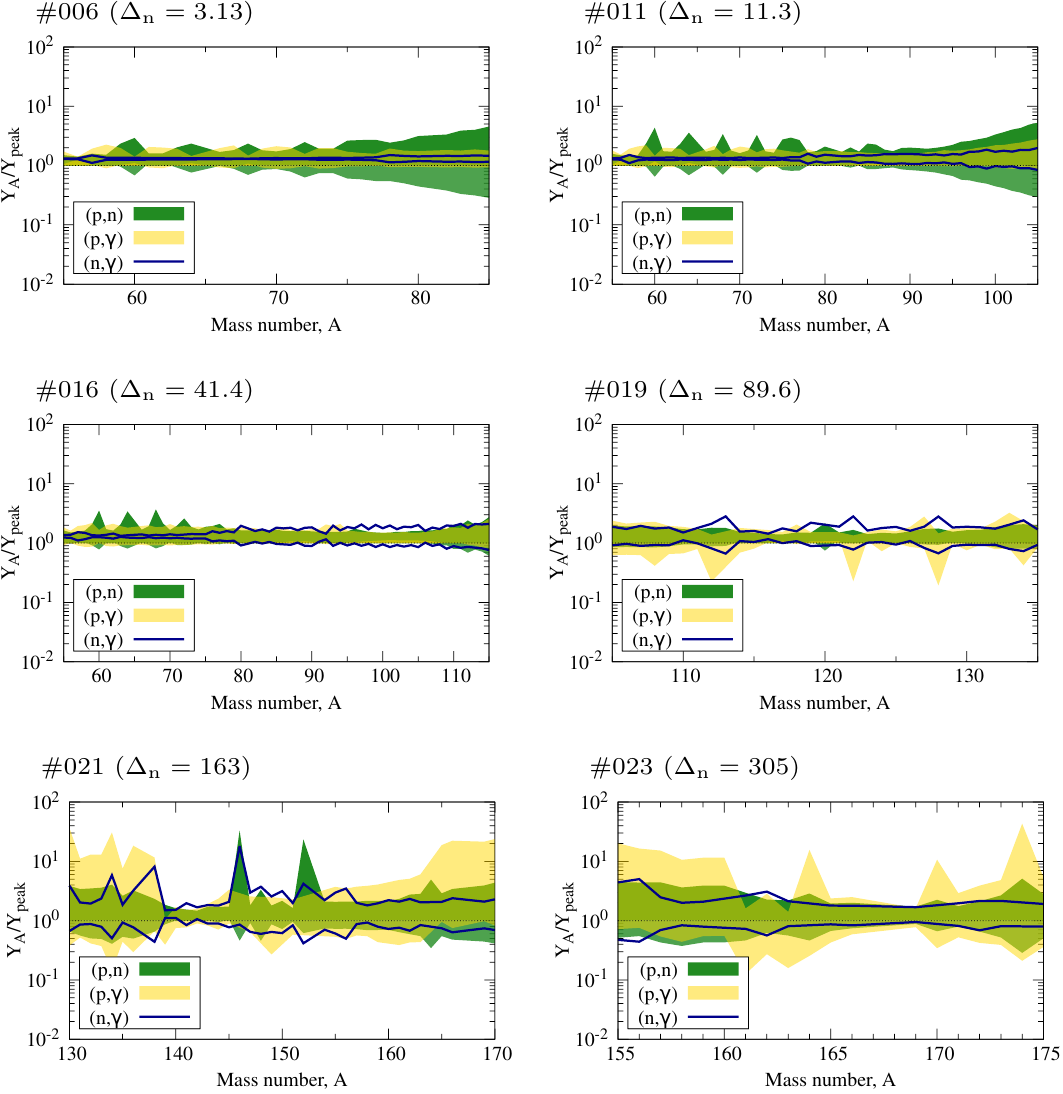}
\caption{\label{fig:keylevelssingle}Final uncertainties obtained in six selected trajectories when only varying (p,n)$\leftrightarrow$(n,p), (n,$\gamma$)$\leftrightarrow$($\gamma$,n), or (p,$\gamma$)$\leftrightarrow$($\gamma$,p) reactions, respectively. Note that the mass number ranges are different in the different panels.}
\end{figure*}

Another crucial reaction is $\iso{Ni}{56}$(n,\,p)$\iso{Co}{56}$. It is the first reaction in the path converting the \iso{Ni}{56} seed to heavier nuclides. Therefore it determines the efficiency of the \nup process and all abundances created, regardless of the detailed conditions. Fig.~\ref{fig:ni56np} shows the impact of a variation of the $\iso{Ni}{56}$(n,\,p)$\iso{Co}{56}$ reaction rate on abundances in all trajectories. Similar to the \triplea rate, the resulting abundances are extremely sensitive to this rate. Therefore we do not include this reaction in the further MC rate variations as its uncertainty would cover all other uncertainties. The results presented in Section~\ref{sec:results} were obtained using the $\iso{Ni}{56}$(n,\,p)$\iso{Co}{56}$ rate of \citet{2000ADNDT..75....1R}.

\section{Results and Discussion}
\label{sec:results}

Based on the thermodynamical parameters described in Section~\ref{sec:astrotracers} and given in Table~\ref{tab:conditions}, we performed nucleosynthesis calculations with the nuclear reaction network specified in Section~\ref{sec:montecarlo}. The final mass fractions of nuclei produced in the \nup process for selected trajectories are shown in Fig.~\ref{fig:finalabuns}. For comparison, in Fig.~\ref{fig:finalabuns} the obtained mass fractions are shown for two \triplea rates found in literature (as discussed in Section~\ref{sec:triplea}).

For trajectories \#06, \#11, \#16, \#19, \#21, and \#23, the total uncertainties originating from the combined action of all varied rates are given in Tables \ref{tab:uncertall1} and \ref{tab:uncertall2} and shown in Figs.~\ref{fig:uncertall1} and \ref{fig:uncertall2}, respectively. Only nuclides which are produced with mass fractions larger than $2\times 10^{-5}$ are included in these figures and tables. The ``up'' and ``down'' factors in Tables \ref{tab:uncertall1} and \ref{tab:uncertall2} are to be taken relative to the abundance value $Y_{50}$ (50\% of the cumulative frequency for the $Y$ distribution). They encompass the range of abundance values obtained in 90\% of the MC runs and can be viewed as a 90\% confidence interval. The abundance $Y_\mathrm{peak}$, on the other hand, is the abundance value at the peak of the probability distribution, i.e., the most probable abundance when considering all MC variations. The values of $Y_{50}$ and $Y_\mathrm{peak}$ do not have to coincide because the probability distribution is asymmetric. Especially for very flat distributions, $Y_{50}$ may differ considerably from $Y_\mathrm{peak}$. The probability distribution is visualised by the colour shade for each nuclide in Figs.~\ref{fig:uncertall1} and \ref{fig:uncertall2}. For further details, see Fig.~5 in \citet{2016MNRAS.463.4153R} and Fig.~2 in \citet{2017MNRAS.469.1752N}, and the detailed discussion in Section~2.3 of \citet{2016MNRAS.463.4153R}.

We find generally larger production uncertainties than in our previous studies of other nucleosynthesis processes but still mostly below a factor of three for the trajectories below \#19. The uncertainties become larger in trajectory \#19 and above, eventually reaching factors of about 40 in trajectories \#21 and \#23. The reason for this increase is that these trajectories mainly produce the heavier mass range and the efficiency of the flow towards heavier nuclides is impacted by all the reactions starting from $\iso{Ni}{56}$. Whether or not the heavier nuclides can be produced at all and where the nucleosynthesis path lies is determined by the common action of all reactions in the path. Furthermore, the far end of the nucleosynthesis path is not reached in equilibrium, making individual reactions, and competition between them, more important. In consequence, many reaction uncertainties are convolved, the combined effect strongly ``wagging the tail'' of the path in the heavier mass range. This is also reflected in the fact that no key rates (see below) were found in trajectories \#19--\#23.

Key rates are those rates which dominate the uncertainty of a given nuclide. Key rates identified in all the investigated trajectories are given in Tables \ref{tab:keyrates1}--\ref{tab:keyrates3}. It should be noted that only rates for target nuclides of Fe and above were varied and the \triplea rate and the $\iso{Ni}{56}$(n,\,p)$\iso{Co}{56}$ rate were kept at their chosen standard descriptions for these cases, see Section~\ref{sec:triplea}. For which nuclides key rates appear for a given trajectory mainly depends on how far up to larger mass numbers the reaction flow continues. On the other hand, as can be seen in Fig.~\ref{fig:finalabuns}, trajectories producing heavier nuclides underproduce the lighter mass range. This trend is reflected in the key rate tables, which do not show key rates for lighter nuclides for trajectories producing the heavier mass range. Furthermore, even when nuclides are produced at an appreciable level, not all of them have their uncertainty connected to a single key rate. In this case, several rates contribute to the production uncertainty, with none of them dominating the contribution to the total uncertainty.

As in our previous investigations, key rates were assigned different levels. The most important rates are at level~1. Level~2 key rates are found after removing the level~1 rates from the MC variations. They determine the uncertainty in the production of a given nuclide assuming that the level~1 key rate has been determined. Similarly, level~3 key rates are defined as dominating the abundance uncertainties after level~1 and level~2 key rates have been determined. Fig.~\ref{fig:keylevels} illustrates how the uncertainties are reduced for each key rate level considered. The correlation coefficients for the level~1 key rates (Lv1) are underlined in the Tables~\ref{tab:keyrates1}--\ref{tab:keyrates3}. The \triplea and the $\iso{Ni}{56}$(n,\,p)$\iso{Co}{56}$ rate, excluded from the MC variations, should be considered as level~0 key rates in our scheme, having top priority.

It is not surprising that (n,\,p) rates appear as key rates. They determine the flow into the next isotonic chain and the timescale for proceeding to heavier nuclei. However, also (p,\,$\gamma$)$\leftrightarrow$($\gamma$,\,p) rates are listed in Tables \ref{tab:keyrates1}--\ref{tab:keyrates3}. At first glance, this may appear surprising because a (p,\,$\gamma$)$\leftrightarrow$($\gamma$,\,p) equilibrium is established in the \nup process and in such an equilibrium the abundances do not depend on the individual proton capture or ($\gamma$,\,p) rates. The (p,\,$\gamma$)$\leftrightarrow$($\gamma$,\,p) rates found in the key rate tables, however, are at the edge of the reaction flows, where the rates are slow and either not equilibrated or fall out of equilibrium within our rate variations. Similar conclusions concerning the role of (n,p) reactions and proton captures were also found by \citet{2015EPJWC..9303008F}, varying rates individually.

Neutron captures as key reactions are found in trajectories \#15--\#18. They are competing with (non-equilibrated) proton captures and push the reaction flow further towards stability and towards neutron-rich isotopes.

The impact of only varying (p,n)$\leftrightarrow$(n,p), (n,$\gamma$)$\leftrightarrow$($\gamma$,n), or (p,$\gamma$)$\leftrightarrow$($\gamma$,p) reactions, respectively, is shown in Fig.~\ref{fig:keylevelssingle}. This illustrates the effect of these reaction types in the different mass ranges. We emphasize, however, that only the MC variation of all reaction rates simultaneously provides a realistic assessment of the importance of a rate, as reflected in the definition of the key rates.

The reaction $\iso{Cu}{59}(\mathrm{p},\alpha)\iso{Ni}{56}$ was identified as a level~3 (trajectories \#04, \#07--\#09) or a level~2 key rate (trajectories \#10--\#14) for the abundance of \iso{Fe}{56}, the final decay product of \iso{Ni}{56} after the \nup process has ceased, and for \iso{Ni}{60}. This is part of a reaction cycle as described in Section~\ref{sec:generalfeatures}. A stronger $\iso{Cu}{59}(\mathrm{p},\alpha)\iso{Ni}{56}$ rate cycles material back to \iso{Ni}{56} and hinders the flow to heavier masses \citep{arcofromart}.

A few $\beta^+$ decays were identified as level~3 key rates: \iso{Zn}{58}, \iso{Zn}{59}, and \iso{Ge}{63}. Their uncertainties would only become important after all other (n,\,p) reactions leading out of the respective isotonic chains have been determined.

An overview of all key reactions and how many nuclide abundances are affected by them is given in Table~\ref{tab:keycount}. At the top of the list, which is sorted by the number of reactions with significant impact, are (n,\,p) reactions, as expected.

Trajectories \#07 and higher may contribute to the production of $p$-nuclides (see Section~\ref{sec:intro}). The $p$-nuclides are underlined in Tables \ref{tab:keyrates2} and \ref{tab:keyrates3}. Level 1 key rates concerning $p$ isotopes were only found in trajectories \#15--\#17. For $\iso{Mo}{92,94}$ the key reactions are the proton captures $\iso{Mo}{92}(\mathrm{p},\gamma)\iso{Tc}{93}$ and $\iso{Ru}{94}(\mathrm{p},\gamma)\iso{Rh}{95}$, respectively, indicating that these captures are not in equilibrium under the given conditions. The proton capture on the stable \iso{Mo}{92} was also identified as a key reaction in the $\gamma$-process \citep{2016MNRAS.463.4153R}. In the \nup process it acts at late times, altering the final $\iso{Mo}{92}$ abundance. Regarding the other trajectories, some do not contribute appreciably to the $p$ nuclides and in those which do, the uncertainties of several reactions are combined without a single dominating uncertainty.

The reproduction of the solar $^{92}$Mo/$^{94}$Mo abundance ratio of 1.6 \citep{2003ApJ...591.1220L} in the rp- and \nup processes has been found to be problematic in previous studies \citep[see, e.g.,][]{2004ApJS..151...75W,2009ApJ...690L.135F,2011ApJ...729...46W,XING2018358}. The abundance ratios of possible progenitor nuclides of these Mo isotopes within an isotonic chain are given mainly by the proton separation energies and therefore the attention in previous studies was focused on accurate mass determinations to tackle this problem. Masses are not varied in the present MC study. We find, nevertheless, that also uncertainties in the reaction rates affect not only the individual abundances of $^{92}$Mo and $^{94}$Mo but also their production ratio. This is because a leakage from an equilibrated (p,$\gamma$)$\leftrightarrow$($\gamma$,p) chain can occur depending on the values of proton capture and (n,p) rates. Another reason is that the (p,$\gamma$)$\leftrightarrow$($\gamma$,p) equilibrium is not fully upheld in trajectories only barely producing Mo.

Table \ref{tab:ratios} shows the uncertainties in the $^{92}$Mo/$^{94}$Mo abundance ratio for selected trajectories. Although the standard rates do not reproduce the solar ratio, it is located within the 90\% confidence intervals defined by the ``up'' and ``down'' factors in all trajectories. This indicates that it is feasible to reproduce the solar value by adjusting reaction rates without modifying nuclear masses. It should be noted, however, that the most probable abundance values $Y_\mathrm{peak}$ also show the well-known problem of having too much $^{94}$Mo relative to $^{92}$Mo. Among the trajectories discussed here, trajectory \#16 most efficiently produces the mass range of the Mo isotopes (see also Fig.~\ref{fig:finalabuns}).

The rate $\iso{Mo}{92}+\mathrm{p}\leftrightarrow\gamma+\iso{Tc}{93}$ \citep[see,][and references therein, for relevant to the $\gamma$-process]{2016PhRvC..93d5809M}, which has been identified as a key rate for $\iso{Mo}{92}$ production, is also a key rate affecting the $\iso{Mo}{92}/\iso{Mo}{94}$ ratio. The correlation coefficients are $r_{\rm cor} = -0.66$, $-0.67$, $-0.65$, $-0.70$, $-0.74$, $-0.72$, $-0.72$, and $-0.68$ for trajectories \#16, \#17, \#18, \#19, \#20, \#21, \#22, and \#23, respectively. The negative correlation indicates that the proton capture direction is dominating. An increase in the proton capture rate reduces the $\iso{Mo}{92}$ abundance and produces $\iso{Mo}{94}$ through flows via \iso{Tc}{93}. Continuing from \iso{Tc}{93}, two paths to \iso{Mo}{94} are possible, either $\iso{Tc}{93}(\mathrm{p},\gamma)\iso{Nb}{94}(\mathrm{n},\mathrm{p})\iso{Tc}{94}(\mathrm{n},\mathrm{p})\iso{Mo}{94}$ or $\iso{Tc}{93}(\mathrm{n},\gamma)\iso{Tc}{94}(\mathrm{n},\mathrm{p})\iso{Mo}{94}$. The flow via \iso{Nb}{94} dominates in trajectories \#11--\#20. The participating reactions were not identified as level~1 key reactions, though. In addition, the $({\rm n}, \gamma)$ and $({\rm p}, \gamma)$ reactions on $\iso{Ru}{94}$ followed by $\iso{Tc}{93}({\rm p}, \gamma)\iso{Ru}{94}$ can also have a significant impact on the $\iso{Mo}{92}/\iso{Mo}{94}$ ratio by reducing the final $\iso{Mo}{94}$ abundance,\footnote{$\iso{Mo}{94}$ is partially produced by $\iso{Ru}{94}$ after the \nup process via the decay series $\iso{Ru}{94}(\beta^+)\iso{Tc}{94}(\beta^+)\iso{Mo}{94}$, of which half-lives are $3.11 \times 10^3~{\rm s}$ and $1.76 \times 10^4~{\rm s}$, respectively.} although they are not identified as key reactions.

Concerning the production of Kr, Sr, Y, and Zr (see Section~\ref{sec:intro}), uncertainties of a factor of two are found for all stable isotopes of these elements, as seen in Table~\ref{tab:uncertall1}. As for the Mo isotope ratios discussed above, the reproduction of the solar abundances in the Kr-Zr region relative to the Mo region has proved difficult in previous studies of the \nup process \citep[see, e.g.,][]{2011ApJ...729...46W,XING2018358}. Table~\ref{tab:ratios} also shows the abundances of \iso{Sr}{82} and \iso{Kr}{78} relative to \iso{Mo}{94}. The solar value for the latter (0.82) is found, within uncertainties, in trajectory \#19 and higher. The solar value of the ratio including \iso{Sr}{82} (0.54), on the other hand, can only be reproduced (within uncertainties) at conditions around those represented by trajectory \#19. Thus, conditions close to those of trajectory \#19 can possibly simultaneously reproduce the abundance ratios of the Zr, Sr, and Mo isotopes. It has to be noted, however, that the production of these nuclides is only marginal in this trajectory (see Fig.~\ref{fig:finalabuns}). The dominant production would be in the mass range $114 \lesssim A \lesssim 126$ and thus this region would be strongly overproduced relative to the lighter $p$ nuclides.

\begin{table*}
  \centering
\caption{Total production uncertainties for stable nuclides after decay of progenitors made in the $\nu p$ process. The abundance $Y_{\rm peak}$ is the peak value of the final abundance probability distribution from our MC runs. The uncertainty factors shown for variations up and down enclose a 90\% probability interval and are relative to $Y_{50}$. (Trajectories \#06, \#11, and \#16)\label{tab:uncertall1}}
  \begin{tabular}{ccccccccccccc}
    \hline
\multicolumn{1}{c}{} & \multicolumn{4}{c}{(\#06)} & \multicolumn{4}{c}{(\#11)} & \multicolumn{4}{c}{(\#16)}\\
    Nuclide  & Up & Down & $Y_{50}$ & $Y_{\rm peak}$ & Up & Down & $Y_{50}$ & $Y_{\rm peak}$ & Up & Down & $Y_{50}$ & $Y_{\rm peak}$ \\
    \hline
${}^{58}{\rm Ni}$ &   2.07 &  0.750 & $ 1.38 \times 10^{-6}$ & $ 1.64 \times 10^{-6}$&   2.00 &  0.734 & $ 5.13 \times 10^{-7}$ & $ 6.08 \times 10^{-7}$&        &         \\
${}^{60}{\rm Ni}$ &   3.04 &  0.694 & $ 3.71 \times 10^{-5}$ & $ 4.98 \times 10^{-5}$&   5.23 &  0.753 & $ 3.58 \times 10^{-6}$ & $ 5.55 \times 10^{-6}$&   3.73 &  0.671 & $ 2.15 \times 10^{-7}$ & $ 2.89 \times 10^{-7}$ \\
${}^{61}{\rm Ni}$ &   1.74 &  0.707 & $ 6.36 \times 10^{-6}$ & $ 6.75 \times 10^{-6}$&   1.96 &  0.727 & $ 1.52 \times 10^{-6}$ & $ 1.80 \times 10^{-6}$&        &         \\
${}^{62}{\rm Ni}$ &   1.72 &  0.704 & $ 2.68 \times 10^{-6}$ & $ 2.85 \times 10^{-6}$&   1.76 &  0.657 & $ 8.55 \times 10^{-7}$ & $ 9.08 \times 10^{-7}$&        &         \\
${}^{63}{\rm Cu}$ &   2.14 &  0.713 & $ 1.47 \times 10^{-5}$ & $ 1.74 \times 10^{-5}$&   2.28 &  0.685 & $ 2.82 \times 10^{-6}$ & $ 3.35 \times 10^{-6}$&        &         \\
${}^{64}{\rm Zn}$ &   2.45 &  0.734 & $ 4.69 \times 10^{-5}$ & $ 6.31 \times 10^{-5}$&   3.75 &  0.672 & $ 7.23 \times 10^{-6}$ & $ 9.72 \times 10^{-6}$&   3.80 &  0.666 & $ 2.50 \times 10^{-7}$ & $ 3.36 \times 10^{-7}$ \\
${}^{66}{\rm Zn}$ &   1.73 &  0.674 & $ 5.02 \times 10^{-6}$ & $ 5.33 \times 10^{-6}$&   1.84 &  0.798 & $ 1.90 \times 10^{-6}$ & $ 2.26 \times 10^{-6}$&        &         \\
${}^{67}{\rm Zn}$ &   2.22 &  0.670 & $ 7.65 \times 10^{-6}$ & $ 9.08 \times 10^{-6}$&   2.21 &  0.720 & $ 2.57 \times 10^{-6}$ & $ 3.05 \times 10^{-6}$&        &         \\
${}^{68}{\rm Zn}$ &   2.15 &  0.608 & $ 2.81 \times 10^{-5}$ & $ 3.34 \times 10^{-5}$&   4.12 &  0.764 & $ 6.57 \times 10^{-6}$ & $ 1.02 \times 10^{-5}$&   3.94 &  0.649 & $ 2.13 \times 10^{-7}$ & $ 2.86 \times 10^{-7}$ \\
${}^{69}{\rm Ga}$ &   2.03 &  0.696 & $ 6.47 \times 10^{-6}$ & $ 7.68 \times 10^{-6}$&   1.79 &  0.682 & $ 4.89 \times 10^{-6}$ & $ 5.19 \times 10^{-6}$&        &         \\
${}^{71}{\rm Ga}$ &   1.96 &  0.560 & $ 5.16 \times 10^{-6}$ & $ 5.48 \times 10^{-6}$&   2.08 &  0.736 & $ 3.59 \times 10^{-6}$ & $ 4.26 \times 10^{-6}$&        &         \\
${}^{70}{\rm Ge}$ &   1.88 &  0.604 & $ 2.44 \times 10^{-6}$ & $ 2.59 \times 10^{-6}$&   1.81 &  0.687 & $ 2.26 \times 10^{-6}$ & $ 2.40 \times 10^{-6}$&        &         \\
${}^{72}{\rm Ge}$ &   2.38 &  0.535 & $ 8.25 \times 10^{-6}$ & $ 9.79 \times 10^{-6}$&   3.48 &  0.670 & $ 5.13 \times 10^{-6}$ & $ 6.90 \times 10^{-6}$&   2.96 &  0.721 & $ 1.95 \times 10^{-7}$ & $ 2.62 \times 10^{-7}$ \\
${}^{73}{\rm Ge}$ &   1.97 &  0.544 & $ 2.27 \times 10^{-6}$ & $ 2.41 \times 10^{-6}$&   1.76 &  0.671 & $ 3.28 \times 10^{-6}$ & $ 3.49 \times 10^{-6}$&        &         \\
${}^{75}{\rm As}$ &   2.97 &  0.554 & $ 2.23 \times 10^{-6}$ & $ 3.00 \times 10^{-6}$&   2.63 &  0.667 & $ 3.39 \times 10^{-6}$ & $ 4.03 \times 10^{-6}$&        &         \\
${}^{74}{\rm Se}$ &   1.91 &  0.538 & $ 1.13 \times 10^{-6}$ & $ 1.20 \times 10^{-6}$&   1.70 &  0.700 & $ 2.20 \times 10^{-6}$ & $ 2.34 \times 10^{-6}$&        &         \\
${}^{76}{\rm Se}$ &   2.89 &  0.541 & $ 1.94 \times 10^{-6}$ & $ 2.61 \times 10^{-6}$&   3.10 &  0.722 & $ 3.56 \times 10^{-6}$ & $ 4.79 \times 10^{-6}$&   2.21 &  0.652 & $ 2.12 \times 10^{-7}$ & $ 2.52 \times 10^{-7}$ \\
${}^{77}{\rm Se}$ &   2.93 &  0.502 & $ 1.32 \times 10^{-6}$ & $ 1.78 \times 10^{-6}$&   2.47 &  0.671 & $ 4.61 \times 10^{-6}$ & $ 5.48 \times 10^{-6}$&   1.97 &  0.594 & $ 2.88 \times 10^{-7}$ & $ 3.06 \times 10^{-7}$ \\
${}^{79}{\rm Br}$ &   3.08 &  0.464 & $ 3.34 \times 10^{-7}$ & $ 4.49 \times 10^{-7}$&   1.62 &  0.689 & $ 2.86 \times 10^{-6}$ & $ 3.04 \times 10^{-6}$&   1.79 &  0.593 & $ 2.50 \times 10^{-7}$ & $ 2.65 \times 10^{-7}$ \\
${}^{78}{\rm Kr}$ &   2.59 &  0.435 & $ 3.85 \times 10^{-7}$ & $ 4.57 \times 10^{-7}$&   1.70 &  0.682 & $ 2.40 \times 10^{-6}$ & $ 2.55 \times 10^{-6}$&   1.98 &  0.691 & $ 2.09 \times 10^{-7}$ & $ 2.48 \times 10^{-7}$ \\
${}^{80}{\rm Kr}$ &   3.21 &  0.356 & $ 3.88 \times 10^{-7}$ & $ 4.61 \times 10^{-7}$&   2.06 &  0.748 & $ 3.12 \times 10^{-6}$ & $ 3.70 \times 10^{-6}$&   2.18 &  0.658 & $ 4.15 \times 10^{-7}$ & $ 4.92 \times 10^{-7}$ \\
${}^{82}{\rm Kr}$ &&&        &        &   1.60 &  0.710 & $ 2.95 \times 10^{-6}$ & $ 3.13 \times 10^{-6}$&   1.68 &  0.648 & $ 5.31 \times 10^{-7}$ & $ 5.64 \times 10^{-7}$ \\
${}^{83}{\rm Kr}$ &&&        &        &   1.77 &  0.672 & $ 2.63 \times 10^{-6}$ & $ 2.79 \times 10^{-6}$&   1.73 &  0.629 & $ 4.29 \times 10^{-7}$ & $ 4.55 \times 10^{-7}$ \\
${}^{85}{\rm Rb}$ &&&        &        &   2.24 &  0.707 & $ 1.88 \times 10^{-6}$ & $ 2.24 \times 10^{-6}$&   1.80 &  0.614 & $ 3.65 \times 10^{-7}$ & $ 3.87 \times 10^{-7}$ \\
${}^{84}{\rm Sr}$ &&&        &        &   1.62 &  0.711 & $ 2.01 \times 10^{-6}$ & $ 2.14 \times 10^{-6}$&   1.63 &  0.657 & $ 5.98 \times 10^{-7}$ & $ 6.35 \times 10^{-7}$ \\
${}^{86}{\rm Sr}$ &&&        &        &   2.13 &  0.713 & $ 1.93 \times 10^{-6}$ & $ 2.29 \times 10^{-6}$&   1.82 &  0.628 & $ 5.74 \times 10^{-7}$ & $ 6.10 \times 10^{-7}$ \\
${}^{87}{\rm Sr}$ &&&        &        &   1.68 &  0.656 & $ 1.55 \times 10^{-6}$ & $ 1.64 \times 10^{-6}$&   1.73 &  0.634 & $ 5.45 \times 10^{-7}$ & $ 5.79 \times 10^{-7}$ \\
${}^{88}{\rm Sr}$ &&&        &        &   1.64 &  0.660 & $ 1.41 \times 10^{-6}$ & $ 1.50 \times 10^{-6}$&   1.68 &  0.683 & $ 8.52 \times 10^{-7}$ & $ 9.05 \times 10^{-7}$ \\
${}^{89}{\rm Y}$ &&&        &        &   1.99 &  0.695 & $ 1.11 \times 10^{-6}$ & $ 1.32 \times 10^{-6}$&   1.92 &  0.717 & $ 5.38 \times 10^{-7}$ & $ 6.39 \times 10^{-7}$ \\
${}^{90}{\rm Zr}$ &&&        &        &   2.04 &  0.676 & $ 1.28 \times 10^{-6}$ & $ 1.52 \times 10^{-6}$&   1.99 &  0.721 & $ 8.36 \times 10^{-7}$ & $ 9.93 \times 10^{-7}$ \\
${}^{91}{\rm Zr}$ &&&        &        &   1.92 &  0.659 & $ 1.11 \times 10^{-6}$ & $ 1.31 \times 10^{-6}$&   1.65 &  0.700 & $ 1.38 \times 10^{-6}$ & $ 1.47 \times 10^{-6}$ \\
${}^{92}{\rm Nb}$ &&&&&        &        &        &        &   2.54 &  0.491 & $ 2.38 \times 10^{-7}$ & $ 2.83 \times 10^{-7}$ \\
${}^{93}{\rm Nb}$ &&&        &        &   1.68 &  0.469 & $ 7.74 \times 10^{-7}$ & $ 7.44 \times 10^{-7}$&   2.45 &  0.809 & $ 7.94 \times 10^{-7}$ & $ 1.07 \times 10^{-6}$ \\
${}^{92}{\rm Mo}$ &&&        &        &   1.76 &  0.556 & $ 1.01 \times 10^{-6}$ & $ 1.07 \times 10^{-6}$&   2.11 &  0.740 & $ 1.59 \times 10^{-6}$ & $ 1.89 \times 10^{-6}$ \\
${}^{94}{\rm Mo}$ &&&        &        &   2.15 &  0.550 & $ 7.59 \times 10^{-7}$ & $ 9.01 \times 10^{-7}$&   2.11 &  0.730 & $ 1.91 \times 10^{-6}$ & $ 2.27 \times 10^{-6}$ \\
${}^{95}{\rm Mo}$ &&&        &        &   2.30 &  0.499 & $ 5.14 \times 10^{-7}$ & $ 6.10 \times 10^{-7}$&   1.96 &  0.722 & $ 1.43 \times 10^{-6}$ & $ 1.70 \times 10^{-6}$ \\
${}^{96}{\rm Mo}$ &&&&&        &        &        &        &   3.12 &  0.546 & $ 1.80 \times 10^{-7}$ & $ 2.43 \times 10^{-7}$ \\
${}^{97}{\rm Tc}$ &&&        &        &   3.03 &  0.435 & $ 2.46 \times 10^{-7}$ & $ 3.30 \times 10^{-7}$&   2.05 &  0.690 & $ 1.86 \times 10^{-6}$ & $ 2.21 \times 10^{-6}$ \\
${}^{96}{\rm Ru}$ &&&        &        &   2.51 &  0.417 & $ 8.55 \times 10^{-7}$ & $ 1.02 \times 10^{-6}$&   1.81 &  0.659 & $ 4.54 \times 10^{-6}$ & $ 4.82 \times 10^{-6}$ \\
${}^{98}{\rm Ru}$ &&&        &        &   3.78 &  0.465 & $ 1.31 \times 10^{-7}$ & $ 2.03 \times 10^{-7}$&   1.60 &  0.680 & $ 3.50 \times 10^{-6}$ & $ 3.72 \times 10^{-6}$ \\
${}^{99}{\rm Ru}$ &&&&&        &        &        &        &   1.79 &  0.641 & $ 1.93 \times 10^{-6}$ & $ 2.05 \times 10^{-6}$ \\
${}^{100}{\rm Ru}$ &&&&&        &        &        &        &   1.58 &  0.708 & $ 3.56 \times 10^{-6}$ & $ 3.78 \times 10^{-6}$ \\
${}^{101}{\rm Ru}$ &&&&&        &        &        &        &   1.94 &  0.726 & $ 1.83 \times 10^{-6}$ & $ 2.17 \times 10^{-6}$ \\
${}^{103}{\rm Rh}$ &&&&&        &        &        &        &   1.90 &  0.731 & $ 1.23 \times 10^{-6}$ & $ 1.46 \times 10^{-6}$ \\
${}^{102}{\rm Pd}$ &&&&&        &        &        &        &   1.55 &  0.702 & $ 2.42 \times 10^{-6}$ & $ 2.57 \times 10^{-6}$ \\
${}^{104}{\rm Pd}$ &&&&&        &        &        &        &   1.68 &  0.669 & $ 1.64 \times 10^{-6}$ & $ 1.75 \times 10^{-6}$ \\
${}^{105}{\rm Pd}$ &&&&&        &        &        &        &   2.00 &  0.714 & $ 8.51 \times 10^{-7}$ & $ 1.01 \times 10^{-6}$ \\
${}^{106}{\rm Pd}$ &&&&&        &        &        &        &   2.24 &  0.660 & $ 3.32 \times 10^{-7}$ & $ 3.94 \times 10^{-7}$ \\
${}^{107}{\rm Ag}$ &&&&&        &        &        &        &   1.69 &  0.629 & $ 8.97 \times 10^{-7}$ & $ 9.53 \times 10^{-7}$ \\
${}^{109}{\rm Ag}$ &&&&&        &        &        &        &   2.22 &  0.578 & $ 4.92 \times 10^{-7}$ & $ 5.84 \times 10^{-7}$ \\
${}^{106}{\rm Cd}$ &&&&&        &        &        &        &   1.60 &  0.627 & $ 1.53 \times 10^{-6}$ & $ 1.62 \times 10^{-6}$ \\
${}^{108}{\rm Cd}$ &&&&&        &        &        &        &   1.73 &  0.554 & $ 1.13 \times 10^{-6}$ & $ 1.20 \times 10^{-6}$ \\
${}^{110}{\rm Cd}$ &&&&&        &        &        &        &   2.36 &  0.511 & $ 4.57 \times 10^{-7}$ & $ 5.42 \times 10^{-7}$ \\
${}^{111}{\rm Cd}$ &&&&&        &        &        &        &   2.40 &  0.505 & $ 2.67 \times 10^{-7}$ & $ 3.17 \times 10^{-7}$ \\
${}^{113}{\rm In}$ &&&&&        &        &        &        &   2.88 &  0.513 & $ 2.49 \times 10^{-7}$ & $ 3.35 \times 10^{-7}$ \\
${}^{112}{\rm Sn}$ &&&&&        &        &        &        &   3.07 &  0.414 & $ 3.55 \times 10^{-7}$ & $ 4.78 \times 10^{-7}$ \\
${}^{114}{\rm Sn}$ &&&&&        &        &        &        &   3.82 &  0.497 & $ 1.39 \times 10^{-7}$ & $ 2.16 \times 10^{-7}$ \\
    \hline
  \end{tabular}
\end{table*}

\begin{table*}
  \centering
  \caption{Total production uncertainties for stable nuclides after decay of progenitors made in the $\nu p$ process. The uncertainty factors shown for variations up and down enclose a 90\% probability interval. (Trajectories \#19, \#21, and \#23)\label{tab:uncertall2}}
  \begin{tabular}{ccccccccccccc}
    \hline
\multicolumn{1}{c}{} & \multicolumn{4}{c}{(\#19)} & \multicolumn{4}{c}{(\#21)} & \multicolumn{4}{c}{(\#23)}\\
    Nuclide  & Up & Down & $Y_{50}$ & $Y_{\rm peak}$ & Up & Down & $Y_{50}$ & $Y_{\rm peak}$ & Up & Down & $Y_{50}$ & $Y_{\rm peak}$ \\
    \hline
${}^{104}{\rm Pd}$ &   6.85 &  0.441 & $ 7.69 \times 10^{-8}$ & $ 1.72 \times 10^{-7}$&        &        &        &         \\
${}^{106}{\rm Pd}$ &   3.87 &  0.433 & $ 1.25 \times 10^{-7}$ & $ 1.93 \times 10^{-7}$&        &        &        &         \\
${}^{109}{\rm Ag}$ &   4.24 &  0.444 & $ 2.04 \times 10^{-7}$ & $ 3.73 \times 10^{-7}$&        &        &        &         \\
${}^{108}{\rm Cd}$ &   8.30 &  0.340 & $ 1.35 \times 10^{-7}$ & $ 3.89 \times 10^{-7}$&        &        &        &         \\
${}^{110}{\rm Cd}$ &   1.70 &  0.197 & $ 7.59 \times 10^{-7}$ & $ 6.66 \times 10^{-7}$&        &        &        &         \\
${}^{111}{\rm Cd}$ &   1.66 &  0.262 & $ 3.77 \times 10^{-7}$ & $ 3.31 \times 10^{-7}$&        &        &        &         \\
${}^{113}{\rm In}$ &   10.9 &  0.345 & $ 5.13 \times 10^{-8}$ & $ 1.48 \times 10^{-7}$&        &        &        &         \\
${}^{112}{\rm Sn}$ &   12.4 &  0.484 & $ 1.08 \times 10^{-7}$ & $ 4.34 \times 10^{-7}$&        &        &        &         \\
${}^{114}{\rm Sn}$ &   1.66 &  0.300 & $ 1.82 \times 10^{-6}$ & $ 1.75 \times 10^{-6}$&        &        &        &         \\
${}^{115}{\rm Sn}$ &   1.52 &  0.357 & $ 8.24 \times 10^{-7}$ & $ 7.92 \times 10^{-7}$&        &        &        &         \\
${}^{116}{\rm Sn}$ &   1.33 &  0.631 & $ 2.39 \times 10^{-6}$ & $ 2.30 \times 10^{-6}$&        &        &        &         \\
${}^{117}{\rm Sn}$ &   1.61 &  0.689 & $ 8.08 \times 10^{-7}$ & $ 8.58 \times 10^{-7}$&        &        &        &         \\
${}^{118}{\rm Sn}$ &   2.16 &  0.649 & $ 1.11 \times 10^{-6}$ & $ 1.32 \times 10^{-6}$&        &        &        &         \\
${}^{119}{\rm Sn}$ &   2.09 &  0.703 & $ 5.94 \times 10^{-7}$ & $ 7.05 \times 10^{-7}$&        &        &        &         \\
${}^{121}{\rm Sb}$ &   1.75 &  0.584 & $ 8.72 \times 10^{-7}$ & $ 9.26 \times 10^{-7}$&        &        &        &         \\
${}^{123}{\rm Sb}$ &   1.46 &  0.708 & $ 1.25 \times 10^{-6}$ & $ 1.32 \times 10^{-6}$&        &        &        &         \\
${}^{122}{\rm Te}$ &   9.04 &  0.417 & $ 1.10 \times 10^{-7}$ & $ 3.17 \times 10^{-7}$&        &        &        &         \\
${}^{124}{\rm Te}$ &   1.58 &  0.328 & $ 1.23 \times 10^{-6}$ & $ 1.18 \times 10^{-6}$&        &        &        &         \\
${}^{125}{\rm Te}$ &   1.72 &  0.638 & $ 1.09 \times 10^{-6}$ & $ 1.16 \times 10^{-6}$&        &        &        &         \\
${}^{126}{\rm Te}$ &   1.50 &  0.449 & $ 1.69 \times 10^{-6}$ & $ 1.62 \times 10^{-6}$&        &        &        &         \\
${}^{127}{\rm I}$ &   3.44 &  0.588 & $ 5.37 \times 10^{-7}$ & $ 8.33 \times 10^{-7}$&        &        &        &         \\
${}^{129}{\rm Xe}$ &   2.26 &  0.647 & $ 4.24 \times 10^{-7}$ & $ 5.03 \times 10^{-7}$&        &        &        &         \\
${}^{130}{\rm Xe}$ &   1.98 &  0.379 & $ 5.19 \times 10^{-7}$ & $ 5.52 \times 10^{-7}$&        &        &        &         \\
${}^{131}{\rm Xe}$ &   2.17 &  0.567 & $ 5.74 \times 10^{-7}$ & $ 6.81 \times 10^{-7}$&        &        &        &         \\
${}^{132}{\rm Xe}$ &   2.59 &  0.407 & $ 5.68 \times 10^{-7}$ & $ 6.75 \times 10^{-7}$&   37.3 &  0.577 & $ 1.65 \times 10^{-8}$ & $ 1.10 \times 10^{-7}$&        &         \\
${}^{133}{\rm Cs}$ &   5.94 &  0.604 & $ 1.69 \times 10^{-7}$ & $ 3.10 \times 10^{-7}$&   37.8 &  0.546 & $ 1.86 \times 10^{-8}$ & $ 1.24 \times 10^{-7}$&        &         \\
${}^{135}{\rm Ba}$ &   2.38 &  0.476 & $ 2.36 \times 10^{-7}$ & $ 2.81 \times 10^{-7}$&   19.5 &  0.414 & $ 5.08 \times 10^{-8}$ & $ 3.40 \times 10^{-7}$&        &         \\
${}^{136}{\rm Ba}$ &   2.37 &  0.352 & $ 3.65 \times 10^{-7}$ & $ 3.87 \times 10^{-7}$&        &        &        &         \\
${}^{139}{\rm La}$ &   4.61 &  0.338 & $ 8.25 \times 10^{-8}$ & $ 1.28 \times 10^{-7}$&   1.96 &  0.240 & $ 1.05 \times 10^{-6}$ & $ 1.12 \times 10^{-6}$&        &         \\
${}^{140}{\rm Ce}$ &&&        &        &   1.62 &  0.319 & $ 1.63 \times 10^{-6}$ & $ 1.57 \times 10^{-6}$&        &         \\
${}^{142}{\rm Ce}$ &&&        &        &   1.52 &  0.390 & $ 5.89 \times 10^{-7}$ & $ 5.66 \times 10^{-7}$&        &         \\
${}^{141}{\rm Pr}$ &&&        &        &   1.71 &  0.294 & $ 5.22 \times 10^{-7}$ & $ 5.02 \times 10^{-7}$&        &         \\
${}^{143}{\rm Nd}$ &&&        &        &   2.71 &  0.525 & $ 4.04 \times 10^{-7}$ & $ 5.44 \times 10^{-7}$&        &         \\
${}^{144}{\rm Nd}$ &&&        &        &   3.00 &  0.557 & $ 3.11 \times 10^{-7}$ & $ 4.18 \times 10^{-7}$&        &         \\
${}^{145}{\rm Nd}$ &&&        &        &   4.33 &  0.536 & $ 1.80 \times 10^{-7}$ & $ 2.42 \times 10^{-7}$&        &         \\
${}^{147}{\rm Sm}$ &&&        &        &   5.34 &  0.651 & $ 1.45 \times 10^{-7}$ & $ 2.66 \times 10^{-7}$&        &         \\
${}^{149}{\rm Sm}$ &&&        &        &   13.1 &  0.808 & $ 3.80 \times 10^{-8}$ & $ 1.53 \times 10^{-7}$&        &         \\
${}^{151}{\rm Eu}$ &&&        &        &   4.02 &  0.236 & $ 2.04 \times 10^{-7}$ & $ 3.16 \times 10^{-7}$&        &         \\
${}^{155}{\rm Gd}$ &&&        &        &   7.25 &  0.408 & $ 5.82 \times 10^{-8}$ & $ 1.07 \times 10^{-7}$&        &         \\
${}^{157}{\rm Gd}$ &&&        &        &   10.3 &  0.652 & $ 6.34 \times 10^{-8}$ & $ 1.83 \times 10^{-7}$&        &         \\
${}^{158}{\rm Gd}$ &&&        &        &   8.19 &  0.424 & $ 1.05 \times 10^{-7}$ & $ 2.36 \times 10^{-7}$&   47.2 &  0.465 & $ 1.83 \times 10^{-8}$ & $ 1.23 \times 10^{-7}$ \\
${}^{159}{\rm Tb}$ &&&        &        &   8.73 &  0.376 & $ 8.21 \times 10^{-8}$ & $ 1.84 \times 10^{-7}$&   41.8 &  0.826 & $ 1.76 \times 10^{-8}$ & $ 1.18 \times 10^{-7}$ \\
${}^{161}{\rm Dy}$ &&&&&        &        &        &        &   30.9 &  0.585 & $ 1.68 \times 10^{-8}$ & $ 1.12 \times 10^{-7}$ \\
${}^{162}{\rm Dy}$ &&&&&        &        &        &        &   28.0 &  0.456 & $ 2.48 \times 10^{-8}$ & $ 1.66 \times 10^{-7}$ \\
${}^{163}{\rm Dy}$ &&&&&        &        &        &        &   23.6 &  0.374 & $ 4.99 \times 10^{-8}$ & $ 3.34 \times 10^{-7}$ \\
${}^{165}{\rm Ho}$ &&&&&        &        &        &        &   17.0 &  0.579 & $ 1.18 \times 10^{-7}$ & $ 7.91 \times 10^{-7}$ \\
${}^{166}{\rm Er}$ &&&&&        &        &        &        &   4.33 &  0.245 & $ 2.95 \times 10^{-7}$ & $ 5.40 \times 10^{-7}$ \\
${}^{169}{\rm Tm}$ &&&&&        &        &        &        &   1.87 &  0.311 & $ 8.42 \times 10^{-7}$ & $ 8.94 \times 10^{-7}$ \\
${}^{171}{\rm Yb}$ &&&&&        &        &        &        &   4.31 &  0.346 & $ 2.25 \times 10^{-7}$ & $ 3.49 \times 10^{-7}$ \\
${}^{172}{\rm Yb}$ &&&&&        &        &        &        &   14.9 &  0.543 & $ 4.41 \times 10^{-8}$ & $ 1.77 \times 10^{-7}$ \\
${}^{173}{\rm Yb}$ &&&&&        &        &        &        &   19.4 &  0.569 & $ 4.01 \times 10^{-8}$ & $ 1.61 \times 10^{-7}$ \\
${}^{175}{\rm Lu}$ &&&&&        &        &        &        &   32.9 &  0.671 & $ 3.16 \times 10^{-8}$ & $ 2.11 \times 10^{-7}$ \\
    \hline
  \end{tabular}
\end{table*}

\begin{table*}
\centering
\caption{Key rates dominating the uncertainties for stable nuclides after decay of progenitors made in the $\nu p$ process for trajectories \#01--\#06 and their correlation coefficients $r_\mathrm{cor}$. The correlation factors for the level~1 key rate (Lv1) are underlined.\label{tab:keyrates1}}
\begin{tabular}{lcrrrrrr}
\hline
Nucleus & Reaction  & \#01 & \#02 & \#03 & \#04 & \#05 & \#06 \\
\hline
$               {}^{56}{\rm Fe}$ & ${}^{57}{\rm{Co}} + {\rm p} \leftrightarrow {\rm n} + {}^{57}{\rm{Ni}}$ &       &       &       &       &  0.67 (Lv3) &       \\
$               {}^{56}{\rm Fe}$ & ${}^{56}{\rm{Ni}} + \alpha \leftrightarrow {\rm p} + {}^{59}{\rm{Cu}}$  &       &       &       &  0.78 (Lv3) &       &       \\
$               {}^{57}{\rm Fe}$ & ${}^{56}{\rm{Ni}} + {\rm p} \leftrightarrow \gamma + {}^{57}{\rm{Cu}}$  &  0.65 (Lv3) &       &       &       &       &       \\
$               {}^{57}{\rm Fe}$ & ${}^{57}{\rm{Ni}} + {\rm p} \leftrightarrow \gamma + {}^{58}{\rm{Cu}}$  & -0.67 (Lv3) & \underline{-0.65 (Lv1)} & -0.75 (Lv2) & -0.74 (Lv2) & -0.73 (Lv2) & \underline{-0.65 (Lv1)} \\
$               {}^{59}{\rm Co}$ & ${}^{59}{\rm{Zn}} $($\beta^+$)$^{59}{\rm{Cu}}$  & -0.94 (Lv3) &       &       &
-0.92 (Lv3) & -0.90 (Lv3) & -0.88 (Lv3) \\
$               {}^{59}{\rm Co}$ & ${}^{59}{\rm{Cu}} + {\rm p} \leftrightarrow \gamma + {}^{60}{\rm{Zn}}$  &       &       &       & -0.70 (Lv2) & -0.73 (Lv2) & -0.75 (Lv2) \\
$               {}^{59}{\rm Co}$ & ${}^{59}{\rm{Cu}} + {\rm p} \leftrightarrow {\rm n} + {}^{59}{\rm{Zn}}$  &       &       &       & \underline{-0.67 (Lv1)} & \underline{-0.67 (Lv1)} & \underline{-0.68 (Lv1)} \\
$               {}^{58}{\rm Ni}$ & ${}^{58}{\rm{Zn}} $($\beta^+$)$^{58}{\rm{Cu}}$  & -0.72 (Lv3) &
-0.69 (Lv3) &       &       &       &       \\
$               {}^{58}{\rm Ni}$ & ${}^{57}{\rm{Cu}} + {\rm p} \leftrightarrow \gamma + {}^{58}{\rm{Zn}}$  &  0.69 (Lv2) &  0.69 (Lv2) &       &       &       &       \\
$               {}^{58}{\rm Ni}$ & ${}^{58}{\rm{Cu}} + {\rm p} \leftrightarrow \gamma + {}^{59}{\rm{Zn}}$  & \underline{-0.67 (Lv1)} & \underline{-0.75 (Lv1)} & \underline{-0.79 (Lv1)} & \underline{-0.78 (Lv1)} & \underline{-0.77 (Lv1)} & \underline{-0.77 (Lv1)} \\
$               {}^{60}{\rm Ni}$ & ${}^{59}{\rm{Cu}} + {\rm p} \leftrightarrow \gamma + {}^{60}{\rm{Zn}}$  &  0.67 (Lv2) &       &       &       &       &       \\
$               {}^{60}{\rm Ni}$ & ${}^{57}{\rm{Co}} + {\rm p} \leftrightarrow {\rm n} + {}^{57}{\rm{Ni}}$  &       & -0.65 (Lv3) & -0.68 (Lv2) & -0.66 (Lv2) & -0.70 (Lv3) &       \\
$               {}^{60}{\rm Ni}$ & ${}^{56}{\rm{Ni}} + \alpha \leftrightarrow {\rm p} + {}^{59}{\rm{Cu}}$  &       &       &       & -0.66 (Lv3) &       &       \\
$               {}^{60}{\rm Ni}$ & ${}^{60}{\rm{Cu}} + {\rm p} \leftrightarrow {\rm n} + {}^{60}{\rm{Zn}}$  &       & \underline{-0.74 (Lv1)} & \underline{-0.83 (Lv1)} & \underline{-0.87 (Lv1)} & \underline{-0.88 (Lv1)} & \underline{-0.88 (Lv1)} \\
$               {}^{61}{\rm Ni}$ & ${}^{60}{\rm{Cu}} + {\rm p} \leftrightarrow \gamma + {}^{61}{\rm{Zn}}$  &  0.78 (Lv3) &  0.75 (Lv2) &  0.72 (Lv2) &  0.69 (Lv2) &  0.68 (Lv2) &  0.66 (Lv2) \\
$               {}^{61}{\rm Ni}$ & ${}^{60}{\rm{Zn}} + {\rm p} \leftrightarrow \gamma + {}^{61}{\rm{Ga}}$  &  0.67 (Lv2) &       &       &       &       &       \\
$               {}^{61}{\rm Ni}$ & ${}^{61}{\rm{Zn}} + {\rm p} \leftrightarrow \gamma + {}^{62}{\rm{Ga}}$  & \underline{-0.65 (Lv1)} & \underline{-0.74 (Lv1)} & \underline{-0.78 (Lv1)} & \underline{-0.77 (Lv1)} & \underline{-0.77 (Lv1)} & \underline{-0.77 (Lv1)} \\
$               {}^{62}{\rm Ni}$ & ${}^{62}{\rm{Zn}} + {\rm p} \leftrightarrow \gamma + {}^{63}{\rm{Ga}}$  & -0.80 (Lv3) & -0.87 (Lv3) & -0.90 (Lv3) & -0.65 (Lv3) &       & -0.66 (Lv3) \\
$               {}^{62}{\rm Ni}$ & ${}^{62}{\rm{Ga}} + {\rm p} \leftrightarrow \gamma + {}^{63}{\rm{Ge}}$  & -0.71 (Lv2) & -0.69 (Lv2) & -0.65 (Lv2) & -0.66 (Lv3) &       &       \\
$               {}^{63}{\rm Cu}$ & ${}^{63}{\rm{Ge}} $($\beta^+$)$^{63}{\rm{Ga}}$  & -0.82 (Lv3) &
-0.75 (Lv3) &       &       &       &       \\
$               {}^{63}{\rm Cu}$ & ${}^{63}{\rm{Ga}} + {\rm p} \leftrightarrow \gamma + {}^{64}{\rm{Ge}}$  & -0.71 (Lv2) & -0.71 (Lv2) &       &       &       &       \\
$               {}^{63}{\rm Cu}$ & ${}^{60}{\rm{Cu}} + {\rm p} \leftrightarrow {\rm n} + {}^{60}{\rm{Zn}}$  & \underline{ 0.73 (Lv1)} & \underline{ 0.67 (Lv1)} &       &       &       &       \\
$               {}^{64}{\rm Zn}$ & ${}^{60}{\rm{Cu}} + {\rm p} \leftrightarrow {\rm n} + {}^{60}{\rm{Zn}}$  & \underline{ 0.90 (Lv1)} & \underline{ 0.88 (Lv1)} & \underline{ 0.69 (Lv1)} &       &       &       \\
$               {}^{64}{\rm Zn}$ & ${}^{64}{\rm{Ga}} + {\rm p} \leftrightarrow {\rm n} + {}^{64}{\rm{Ge}}$  &       &       &       & \underline{-0.69 (Lv1)} & \underline{-0.75 (Lv1)} & \underline{-0.79 (Lv1)} \\
$               {}^{67}{\rm Zn}$ & ${}^{67}{\rm{As}} + {\rm p} \leftrightarrow \gamma + {}^{68}{\rm{Se}}$  & -0.69 (Lv2) & -0.72 (Lv2) & -0.78 (Lv2) & -0.77 (Lv2) & -0.75 (Lv2) & \underline{-0.65 (Lv1)} \\
$               {}^{68}{\rm Zn}$ & ${}^{64}{\rm{Ga}} + {\rm p} \leftrightarrow {\rm n} + {}^{64}{\rm{Ge}}$  & \underline{ 0.77 (Lv1)} & \underline{ 0.74 (Lv1)} & \underline{ 0.73 (Lv1)} &       &       &       \\
$               {}^{68}{\rm Zn}$ & ${}^{68}{\rm{As}} + {\rm p} \leftrightarrow {\rm n} + {}^{68}{\rm{Se}}$  &       &       &       & -0.78 (Lv2) & -0.83 (Lv2) & \underline{-0.70 (Lv1)} \\
$               {}^{69}{\rm Ga}$ & ${}^{69}{\rm{Se}} + {\rm p} \leftrightarrow \gamma + {}^{70}{\rm{Br}}$  &       &       & -0.68 (Lv3) & -0.74 (Lv3) & -0.75 (Lv3) & -0.73 (Lv2) \\
$               {}^{69}{\rm Ga}$ & ${}^{68}{\rm{As}} + {\rm p} \leftrightarrow {\rm n} + {}^{68}{\rm{Se}}$  &  0.67 (Lv2) &  0.65 (Lv3) &  0.65 (Lv2) &       &       &       \\
$               {}^{71}{\rm Ga}$ & ${}^{71}{\rm{Br}} + {\rm p} \leftrightarrow \gamma + {}^{72}{\rm{Kr}}$  &       &       &       & -0.70 (Lv3) & -0.71 (Lv3) & -0.73 (Lv2) \\
$               {}^{71}{\rm Ga}$ & ${}^{68}{\rm{As}} + {\rm p} \leftrightarrow {\rm n} + {}^{68}{\rm{Se}}$  &  0.66 (Lv2) &       &       &       &       &       \\
$               {}^{70}{\rm Ge}$ & ${}^{70}{\rm{Se}} + {\rm p} \leftrightarrow \gamma + {}^{71}{\rm{Br}}$  &       &       &       &       & -0.65 (Lv3) & -0.68 (Lv2) \\
$               {}^{70}{\rm Ge}$ & ${}^{70}{\rm{Br}} + {\rm p} \leftrightarrow \gamma + {}^{71}{\rm{Kr}}$  &       &       &       &       &       & -0.71 (Lv3) \\
$               {}^{72}{\rm Ge}$ & ${}^{68}{\rm{As}} + {\rm p} \leftrightarrow {\rm n} + {}^{68}{\rm{Se}}$  &  0.77 (Lv2) &       &       &       &       &       \\
$               {}^{72}{\rm Ge}$ & ${}^{72}{\rm{Br}} + {\rm p} \leftrightarrow {\rm n} + {}^{72}{\rm{Kr}}$  &       &       &       &       & -0.69 (Lv3) & -0.77 (Lv2) \\
$               {}^{73}{\rm Ge}$ & ${}^{73}{\rm{Kr}} + {\rm p} \leftrightarrow \gamma + {}^{74}{\rm{Rb}}$  &       &       &       &       &       & -0.68 (Lv3) \\
$               {}^{75}{\rm As}$ & ${}^{72}{\rm{Br}} + {\rm p} \leftrightarrow {\rm n} + {}^{72}{\rm{Kr}}$  &  0.67 (Lv3) &       &       &       &       &       \\
$               {}^{75}{\rm As}$ & ${}^{75}{\rm{Rb}} + {\rm p} \leftrightarrow {\rm n} + {}^{75}{\rm{Sr}}$  &       &       &       &       &       & -0.67 (Lv3) \\
\hline
\end{tabular}
\end{table*}

\begin{table*}
\centering
\caption{Same as Table~\ref{tab:keyrates1} but for trajectories \#07--\#12. Underlined nuclides are $p$ nuclides.\label{tab:keyrates2}}
\begin{tabular}{lcrrrrrr}
\hline
Nucleus & Reaction  & \#07 & \#08 & \#09 & \#10 & \#11 & \#12 \\
\hline
$               {}^{56}{\rm Fe}$ & ${}^{59}{\rm{Cu}} + {\rm p} \leftrightarrow \gamma + {}^{60}{\rm{Zn}}$  &       &       &       &       &       & -0.65 (Lv3) \\
$               {}^{56}{\rm Fe}$ & ${}^{56}{\rm{Ni}} + \alpha \leftrightarrow {\rm p} + {}^{59}{\rm{Cu}}$  &  0.66 (Lv3) &  0.69 (Lv3) &  0.66 (Lv3) &  0.66 (Lv2) &  0.67 (Lv2) &  0.68 (Lv2) \\
$               {}^{57}{\rm Fe}$ & ${}^{57}{\rm{Ni}} + {\rm p} \leftrightarrow \gamma + {}^{58}{\rm{Cu}}$  & \underline{-0.66 (Lv1)} & \underline{-0.65 (Lv1)} & -0.66 (Lv2) & -0.70 (Lv3) & -0.70 (Lv3) & -0.69 (Lv3) \\
$               {}^{59}{\rm Co}$ & ${}^{59}{\rm{Zn}} $($\beta^+$)$^{59}{\rm{Cu}}$  & -0.83 (Lv3) &
-0.76 (Lv3) &       &       &       &       \\
$               {}^{59}{\rm Co}$ & ${}^{59}{\rm{Cu}} + {\rm p} \leftrightarrow \gamma + {}^{60}{\rm{Zn}}$  & -0.77 (Lv2) & -0.77 (Lv2) & -0.78 (Lv3) & -0.81 (Lv3) & -0.81 (Lv3) & -0.80 (Lv3) \\
$               {}^{59}{\rm Co}$ & ${}^{59}{\rm{Cu}} + {\rm p} \leftrightarrow {\rm n} + {}^{59}{\rm{Zn}}$  & \underline{-0.67 (Lv1)} & \underline{-0.66 (Lv1)} & -0.66 (Lv2) &       &       &       \\
$               {}^{58}{\rm Ni}$ & ${}^{58}{\rm{Cu}} + {\rm p} \leftrightarrow \gamma + {}^{59}{\rm{Zn}}$  & \underline{-0.75 (Lv1)} & \underline{-0.74 (Lv1)} & \underline{-0.71 (Lv1)} & \underline{-0.68 (Lv1)} & -0.70 (Lv3) & -0.66 (Lv3) \\
$               {}^{60}{\rm Ni}$ & ${}^{59}{\rm{Cu}} + {\rm p} \leftrightarrow {\rm n} + {}^{59}{\rm{Zn}}$  &       &       & -0.75 (Lv2) & -0.78 (Lv2) & -0.74 (Lv2) & -0.68 (Lv2) \\
$               {}^{60}{\rm Ni}$ & ${}^{60}{\rm{Cu}} + {\rm p} \leftrightarrow {\rm n} + {}^{60}{\rm{Zn}}$  & \underline{-0.88 (Lv1)} & \underline{-0.88 (Lv1)} & \underline{-0.87 (Lv1)} & \underline{-0.86 (Lv1)} & \underline{-0.85 (Lv1)} & \underline{-0.84 (Lv1)} \\
$               {}^{61}{\rm Ni}$ & ${}^{60}{\rm{Cu}} + {\rm p} \leftrightarrow \gamma + {}^{61}{\rm{Zn}}$  &  0.66 (Lv2) &  0.66 (Lv2) &       &       &       &       \\
$               {}^{61}{\rm Ni}$ & ${}^{61}{\rm{Zn}} + {\rm p} \leftrightarrow \gamma + {}^{62}{\rm{Ga}}$  & \underline{-0.75 (Lv1)} & \underline{-0.72 (Lv1)} & \underline{-0.69 (Lv1)} & -0.71 (Lv2) & -0.67 (Lv2) & -0.65 (Lv2) \\
$               {}^{62}{\rm Ni}$ & ${}^{62}{\rm{Zn}} + {\rm p} \leftrightarrow \gamma + {}^{63}{\rm{Ga}}$  & -0.67 (Lv2) & -0.68 (Lv2) & -0.69 (Lv3) & -0.70 (Lv3) & -0.70 (Lv3) & -0.69 (Lv3) \\
$               {}^{62}{\rm Ni}$ & ${}^{62}{\rm{Ga}} + {\rm p} \leftrightarrow \gamma + {}^{63}{\rm{Ge}}$  & -0.81 (Lv3) & -0.80 (Lv3) &       &       &       &       \\
$               {}^{63}{\rm Cu}$ & ${}^{63}{\rm{Ga}} + {\rm p} \leftrightarrow \gamma + {}^{64}{\rm{Ge}}$  &       &       & -0.77 (Lv3) & -0.74 (Lv2) & -0.77 (Lv3) & -0.75 (Lv3) \\
$               {}^{63}{\rm Cu}$ & ${}^{63}{\rm{Ga}} + {\rm p} \leftrightarrow {\rm n} + {}^{63}{\rm{Ge}}$  &       & -0.65 (Lv3) & -0.67 (Lv2) & \underline{-0.65 (Lv1)} & -0.67 (Lv2) & -0.65 (Lv2) \\
$               {}^{64}{\rm Zn}$ & ${}^{63}{\rm{Ga}} + {\rm p} \leftrightarrow {\rm n} + {}^{63}{\rm{Ge}}$  &       &       &       &       &       & -0.65 (Lv2) \\
$               {}^{64}{\rm Zn}$ & ${}^{64}{\rm{Ga}} + {\rm p} \leftrightarrow {\rm n} + {}^{64}{\rm{Ge}}$  & \underline{-0.82 (Lv1)} & \underline{-0.84 (Lv1)} & \underline{-0.86 (Lv1)} & \underline{-0.86 (Lv1)} & \underline{-0.85 (Lv1)} & \underline{-0.85 (Lv1)} \\
$               {}^{67}{\rm Zn}$ & ${}^{67}{\rm{As}} + {\rm p} \leftrightarrow \gamma + {}^{68}{\rm{Se}}$  & \underline{-0.66 (Lv1)} & \underline{-0.66 (Lv1)} & \underline{-0.67 (Lv1)} & \underline{-0.67 (Lv1)} & \underline{-0.66 (Lv1)} & -0.67 (Lv3) \\
$               {}^{68}{\rm Zn}$ & ${}^{68}{\rm{As}} + {\rm p} \leftrightarrow {\rm n} + {}^{68}{\rm{Se}}$  & \underline{-0.76 (Lv1)} & \underline{-0.80 (Lv1)} & \underline{-0.82 (Lv1)} & \underline{-0.84 (Lv1)} & \underline{-0.85 (Lv1)} & \underline{-0.84 (Lv1)} \\
$               {}^{69}{\rm Ga}$ & ${}^{69}{\rm{Se}} + {\rm p} \leftrightarrow \gamma + {}^{70}{\rm{Br}}$  & -0.71 (Lv2) & -0.68 (Lv2) &       &       &       &       \\
$               {}^{71}{\rm Ga}$ & ${}^{71}{\rm{Br}} + {\rm p} \leftrightarrow \gamma + {}^{72}{\rm{Kr}}$  & -0.72 (Lv2) & -0.68 (Lv2) & -0.67 (Lv2) & -0.67 (Lv3) &       &       \\
$               {}^{70}{\rm Ge}$ & ${}^{70}{\rm{Se}} + {\rm p} \leftrightarrow \gamma + {}^{71}{\rm{Br}}$  & -0.69 (Lv2) & -0.69 (Lv2) & \underline{-0.65 (Lv1)} & \underline{-0.66 (Lv1)} & \underline{-0.68 (Lv1)} & \underline{-0.70 (Lv1)} \\
$               {}^{70}{\rm Ge}$ & ${}^{70}{\rm{Br}} + {\rm p} \leftrightarrow \gamma + {}^{71}{\rm{Kr}}$  & -0.71 (Lv3) & -0.67 (Lv3) &       &       &       &       \\
$               {}^{72}{\rm Ge}$ & ${}^{72}{\rm{Br}} + {\rm p} \leftrightarrow {\rm n} + {}^{72}{\rm{Kr}}$  & \underline{-0.66 (Lv1)} & \underline{-0.73 (Lv1)} & \underline{-0.77 (Lv1)} & \underline{-0.78 (Lv1)} & \underline{-0.79 (Lv1)} & \underline{-0.79 (Lv1)} \\
$               {}^{73}{\rm Ge}$ & ${}^{73}{\rm{Kr}} + {\rm p} \leftrightarrow \gamma + {}^{74}{\rm{Rb}}$  & -0.68 (Lv2) & -0.69 (Lv2) & -0.65 (Lv2) & -0.65 (Lv3) &       &       \\
$               {}^{75}{\rm As}$ & ${}^{75}{\rm{Rb}} + {\rm p} \leftrightarrow {\rm n} + {}^{75}{\rm{Sr}}$  & -0.72 (Lv2) & -0.75 (Lv2) & \underline{-0.67 (Lv1)} & \underline{-0.68 (Lv1)} & \underline{-0.67 (Lv1)} & \underline{-0.65 (Lv1)} \\
$   \underline{{}^{74}{\rm Se}}$ & ${}^{74}{\rm{Kr}} + {\rm p} \leftrightarrow \gamma + {}^{75}{\rm{Rb}}$  & -0.67 (Lv2) & -0.70 (Lv2) & -0.70 (Lv2) & -0.70 (Lv2) & -0.66 (Lv2) & -0.67 (Lv3) \\
$               {}^{76}{\rm Se}$ & ${}^{76}{\rm{Rb}} + {\rm p} \leftrightarrow {\rm n} + {}^{76}{\rm{Sr}}$  & -0.72 (Lv2) & \underline{-0.67 (Lv1)} & \underline{-0.72 (Lv1)} & \underline{-0.74 (Lv1)} & \underline{-0.73 (Lv1)} & \underline{-0.71 (Lv1)} \\
$               {}^{77}{\rm Se}$ & ${}^{77}{\rm{Rb}} + {\rm p} \leftrightarrow {\rm n} + {}^{77}{\rm{Sr}}$  & -0.69 (Lv3) & -0.75 (Lv2) & \underline{-0.72 (Lv1)} & \underline{-0.75 (Lv1)} & \underline{-0.74 (Lv1)} & \underline{-0.71 (Lv1)} \\
$   \underline{{}^{78}{\rm Kr}}$ & ${}^{78}{\rm{Sr}} + {\rm p} \leftrightarrow \gamma + {}^{79}{\rm{Y}}$  &       &       & -0.66 (Lv3) & -0.65 (Lv2) &       &       \\
$               {}^{80}{\rm Kr}$ & ${}^{80}{\rm{Y}} + {\rm p} \leftrightarrow {\rm n} + {}^{80}{\rm{Zr}}$  &       &       &       & -0.66 (Lv3) &       &       \\
$               {}^{85}{\rm Rb}$ & ${}^{85}{\rm{Nb}} + {\rm p} \leftrightarrow {\rm n} + {}^{85}{\rm{Mo}}$  &       &       &       & -0.65 (Lv3) & -0.67 (Lv2) & -0.65 (Lv3) \\
$               {}^{86}{\rm Sr}$ & ${}^{86}{\rm{Nb}} + {\rm p} \leftrightarrow {\rm n} + {}^{86}{\rm{Mo}}$  &       &       &       &       & -0.66 (Lv3) &       \\
\hline
\end{tabular}
\end{table*}

\begin{table*}
\centering
\caption{Same as Table~\ref{tab:keyrates1} but for trajectories \#13--\#18. Underlined nuclides are $p$
nuclides.\label{tab:keyrates3}}
\begin{tabular}{lcrrrrrr}
\hline
Nucleus & Reaction  & \#13 & \#14 & \#15 & \#16 & \#17 & \#18 \\
\hline
$               {}^{56}{\rm Fe}$ & ${}^{59}{\rm{Cu}} + {\rm p} \leftrightarrow \gamma + {}^{60}{\rm{Zn}}$  & -0.65 (Lv3) &       &       &       &       &       \\
$               {}^{56}{\rm Fe}$ & ${}^{56}{\rm{Ni}} + \alpha \leftrightarrow {\rm p} + {}^{59}{\rm{Cu}}$  &  0.67 (Lv2) &  0.65 (Lv2) &       &       &       &       \\
$               {}^{57}{\rm Fe}$ & ${}^{57}{\rm{Ni}} + {\rm p} \leftrightarrow \gamma + {}^{58}{\rm{Cu}}$  & -0.66 (Lv3) &       &       &       &       &       \\
$               {}^{60}{\rm Ni}$ & ${}^{59}{\rm{Cu}} + {\rm p} \leftrightarrow {\rm n} + {}^{59}{\rm{Zn}}$  & -0.66 (Lv3) &       &       &       &       &       \\
$               {}^{60}{\rm Ni}$ & ${}^{60}{\rm{Cu}} + {\rm p} \leftrightarrow {\rm n} + {}^{60}{\rm{Zn}}$  & \underline{-0.82 (Lv1)} & \underline{-0.81 (Lv1)} & \underline{-0.78 (Lv1)} & \underline{-0.75 (Lv1)} & \underline{-0.69 (Lv1)} &       \\
$               {}^{64}{\rm Zn}$ & ${}^{64}{\rm{Ga}} + {\rm p} \leftrightarrow {\rm n} + {}^{64}{\rm{Ge}}$  & \underline{-0.83 (Lv1)} & \underline{-0.80 (Lv1)} & \underline{-0.75 (Lv1)} & \underline{-0.70 (Lv1)} &       &       \\
$               {}^{68}{\rm Zn}$ & ${}^{68}{\rm{As}} + {\rm p} \leftrightarrow {\rm n} + {}^{68}{\rm{Se}}$  & \underline{-0.84 (Lv1)} & \underline{-0.81 (Lv1)} & \underline{-0.75 (Lv1)} & \underline{-0.68 (Lv1)} &       &       \\
$               {}^{70}{\rm Ge}$ & ${}^{70}{\rm{Se}} + {\rm p} \leftrightarrow \gamma + {}^{71}{\rm{Br}}$  & \underline{-0.70 (Lv1)} & \underline{-0.68 (Lv1)} & -0.65 (Lv2) &       &       &       \\
$               {}^{72}{\rm Ge}$ & ${}^{72}{\rm{Br}} + {\rm p} \leftrightarrow {\rm n} + {}^{72}{\rm{Kr}}$  & \underline{-0.78 (Lv1)} & \underline{-0.75 (Lv1)} & \underline{-0.66 (Lv1)} &       &       &       \\
$               {}^{76}{\rm Se}$ & ${}^{76}{\rm{Rb}} + {\rm p} \leftrightarrow {\rm n} + {}^{76}{\rm{Sr}}$  & \underline{-0.68 (Lv1)} &       &       &       &       &       \\
$               {}^{77}{\rm Se}$ & ${}^{77}{\rm{Rb}} + {\rm p} \leftrightarrow {\rm n} + {}^{77}{\rm{Sr}}$  & \underline{-0.69 (Lv1)} & \underline{-0.65 (Lv1)} &       &       &       &       \\
$               {}^{80}{\rm Kr}$ & ${}^{80}{\rm{Sr}} + {\rm n} \leftrightarrow \gamma + {}^{81}{\rm{Sr}}$  &       &       & -0.65 (Lv2) &       &       &       \\
$               {}^{93}{\rm Nb}$ & ${}^{93}{\rm{Tc}} + {\rm n} \leftrightarrow \gamma + {}^{94}{\rm{Tc}}$  &       &       &       & -0.67 (Lv2) &       &       \\
$               {}^{93}{\rm Nb}$ & ${}^{93}{\rm{Tc}} + {\rm p} \leftrightarrow \gamma + {}^{94}{\rm{Ru}}$  &       &       &       & -0.70 (Lv3) &       &       \\
$   \underline{{}^{92}{\rm Mo}}$ & ${}^{92}{\rm{Mo}} + {\rm p} \leftrightarrow \gamma + {}^{93}{\rm{Tc}}$  &       &       &       & \underline{-0.73 (Lv1)} & \underline{-0.71 (Lv1)} &       \\
$   \underline{{}^{94}{\rm Mo}}$ & ${}^{94}{\rm{Ru}} + {\rm p} \leftrightarrow \gamma + {}^{95}{\rm{Rh}}$  &       & -0.65 (Lv2) & -0.65 (Lv3) & \underline{-0.66 (Lv1)} &       &       \\
$   {}^{97}{\rm Tc}$ & ${}^{97}{\rm{Rh}} + {\rm n} \leftrightarrow \gamma + {}^{98}{\rm{Rh}}$  &       &       &
\underline{-0.70 (Lv1)} & \underline{-0.66 (Lv1)} &       &       \\
$               {}^{99}{\rm Ru}$ & ${}^{99}{\rm{Rh}} + {\rm n} \leftrightarrow \gamma + {}^{100}{\rm{Rh}}$  &       &       &       & -0.65 (Lv3) &       &       \\
$              {}^{100}{\rm Ru}$ & ${}^{100}{\rm{Pd}} + {\rm n} \leftrightarrow \gamma + {}^{101}{\rm{Pd}}$  &       &       &       & -0.66 (Lv2) & \underline{-0.68 (Lv1)} &       \\
$  \underline{{}^{113}{\rm In}}$ & ${}^{113}{\rm{In}} + {\rm n} \leftrightarrow \gamma + {}^{114}{\rm{In}}$  &       &       &       &       & \underline{-0.67 (Lv1)} &       \\
$              {}^{117}{\rm Sn}$ & ${}^{117}{\rm{In}} + {\rm n} \leftrightarrow \gamma + {}^{118}{\rm{In}}$  &       &       &       &       &       & \underline{-0.74 (Lv1)} \\
\hline
\end{tabular}
\end{table*}

\begin{table*}
 \centering
 \caption{Key reaction list sorted by number of affected nuclides per key rate level and by counted number of involved trajectories.\label{tab:keycount}}
        \begin{tabular}{ccccc}
          \hline
          Reaction  &  Level 1 &  Level 2 & Level 3 & Number of trajectories \\
          \hline
\npreac{Cu}{60}{p,n}{Zn}{60}&$\iso{Ni}{60},\iso{Cu}{63},\iso{Zn}{64}$&&&17\\
\npreac{Ga}{64}{p,n}{Ge}{64}&$\iso{Zn}{64},\iso{Zn}{68}$&&&13\\
\npreac{As}{68}{p,n}{Se}{68}&\iso{Zn}{68}&$\iso{Zn}{68},\iso{Ga}{69},\iso{Ga}{71},\iso{Ge}{72}$&\iso{Ga}{69}&16\\
\npreac{Cu}{59}{p,n}{Zn}{59}&\iso{Co}{59}&$\iso{Ni}{60}$,\iso{Co}{59}&\iso{Ni}{60}&10\\
\npreac{Ga}{63}{p,n}{Ge}{63}&\iso{Cu}{63}&$\iso{Cu}{63},\iso{Zn}{64}$&\iso{Cu}{63}&5\\
\npreac{Br}{72}{p,n}{Kr}{72}&\iso{Ge}{72}&\iso{Ge}{72}&$\iso{Ge}{72},\iso{As}{75}$&12\\
\reac{Ni}{57}{p,$\gamma$}{Cu}{58}&\iso{Fe}{57}&\iso{Fe}{57}&\iso{Fe}{57}&13\\
\reac{As}{67}{p,$\gamma$}{Se}{68}&\iso{Zn}{67}&\iso{Zn}{67}&\iso{Zn}{67}&12\\
\reac{Se}{70}{p,$\gamma$}{Br}{71}&\iso{Ge}{70}&\iso{Ge}{70}&\iso{Ge}{70}&11\\
\npreac{Rb}{77}{p,n}{Sr}{77}&\iso{Se}{77}&\iso{Se}{77}&\iso{Se}{77}&8\\
\npreac{Rb}{75}{p,n}{Sr}{75}&\iso{As}{75}&\iso{As}{75}&\iso{As}{75}&7\\
\reac{Ru}{94}{p,$\gamma$}{Rh}{95}&\iso{Mo}{94}&\iso{Mo}{94}&\iso{Mo}{94}&3\\
\reac{Zn}{61}{p,$\gamma$}{Ga}{62}&\iso{Ni}{61}&\iso{Ni}{61}&&12\\
\npreac{Rb}{76}{p,n}{Sr}{76}&\iso{Se}{76}&\iso{Se}{76}&&7\\
\reac{Pd}{100}{n,$\gamma$}{Pd}{101}&\iso{Ru}{100}&\iso{Ru}{100}&&2\\
\reac{Cu}{58}{p,$\gamma$}{Zn}{59}&\iso{Ni}{58}&&\iso{Ni}{58}&12\\
\reac{Mo}{92}{p,$\gamma$}{Tc}{93}&\iso{Mo}{92}&&&2\\
\reac{Rh}{97}{n,$\gamma$}{Rh}{98}&\iso{Tc}{97}&&&2\\
\reac{In}{113}{n,$\gamma$}{In}{114}&\iso{In}{113}&&&1\\
\reac{In}{117}{n,$\gamma$}{In}{118}&\iso{Sn}{117}&&&1\\
\reac{Cu}{59}{p,$\gamma$}{Zn}{60}&&$\iso{Co}{59}$,\iso{Ni}{60}&$\iso{Co}{59}$,\iso{Fe}{56}&11\\
\reac{Cu}{59}{p,$\alpha$}{Ni}{56}&&\iso{Fe}{56}&$\iso{Fe}{56}$,\iso{Ni}{60}&9\\
\npreac{Co}{57}{p,n}{Ni}{57}&&\iso{Ni}{60}&$\iso{Fe}{56}$,\iso{Ni}{60}&4\\
\reac{Zn}{62}{p,$\gamma$}{Ga}{63}&&\iso{Ni}{62}&\iso{Ni}{62}&12\\
\reac{Cu}{60}{p,$\gamma$}{Zn}{61}&&\iso{Ni}{61}&\iso{Ni}{61}&8\\
\reac{Br}{71}{p,$\gamma$}{Kr}{72}&&\iso{Ga}{71}&\iso{Ga}{71}&7\\
\reac{Ga}{62}{p,$\gamma$}{Ge}{63}&&\iso{Ni}{62}&\iso{Ni}{62}&6\\
\reac{Ga}{63}{p,$\gamma$}{Ge}{64}&&\iso{Cu}{63}&\iso{Cu}{63}&6\\
\reac{Se}{69}{p,$\gamma$}{Br}{70}&&\iso{Ga}{69}&\iso{Ga}{69}&6\\
\reac{Kr}{74}{p,$\gamma$}{Rb}{75}&&\iso{Se}{74}&\iso{Se}{74}&6\\
\reac{Kr}{73}{p,$\gamma$}{Rb}{74}&&\iso{Ge}{73}&\iso{Ge}{73}&5\\
\npreac{Nb}{85}{p,n}{Mo}{85}&&\iso{Rb}{85}&\iso{Rb}{85}&3\\
\reac{Sr}{78}{p,$\gamma$}{Y}{79}&&\iso{Kr}{78}&\iso{Kr}{78}&2\\
\reac{Cu}{57}{p,$\gamma$}{Zn}{58}&&\iso{Ni}{58}&&2\\
\reac{Zn}{60}{p,$\gamma$}{Ga}{61}&&\iso{Ni}{61}&&1\\
\reac{Sr}{80}{n,$\gamma$}{Sr}{81}&&\iso{Kr}{80}&&1\\
\reac{Tc}{93}{n,$\gamma$}{Tc}{94}&&\iso{Nb}{93}&&1\\
\reac{Tc}{93}{p,$\gamma$}{Ru}{94}&&\iso{Nb}{93}&&1\\
\reac{Zn}{59}{$\beta^+$}{Cu}{59}&&&\iso{Co}{59}&6\\
\reac{Br}{70}{p,$\gamma$}{Kr}{71}&&&\iso{Ge}{70}&3\\
\reac{Zn}{58}{$\beta^+$}{Cu}{58}&&&\iso{Ni}{58}&2\\
\reac{Ge}{63}{$\beta^+$}{Ga}{63}&&&\iso{Cu}{63}&2\\
\reac{Ni}{56}{p,$\gamma$}{Cu}{57}&&&\iso{Fe}{57}&1\\
\npreac{Y}{80}{p,n}{Zr}{80}&&&\iso{Kr}{80}&1\\
\npreac{Nb}{86}{p,n}{Mo}{86}&&&\iso{Sr}{86}&1\\
\reac{Rh}{99}{n,$\gamma$}{Rh}{100}&&&\iso{Ru}{99}&1\\
          \hline
        \end{tabular}
\end{table*}

\begin{table*}
\caption{Uncertainties of isotopic ratios in selected trajectories, given as uncertainty factors relative to the 50\% cumulative probability. The factors enclose a 90\% probability range. Also shown is the most probable value based on $Y_\mathrm{peak}$. The solar system values are 1.6 for $\iso{Mo}{92}/\iso{Mo}{94}$, 0.54 for $\iso{Sr}{84}/\iso{Mo}{94}$, and 0.82 for $\iso{Kr}{78}/\iso{Mo}{94}$ \citep{2003ApJ...591.1220L}.\label{tab:ratios}}
\begin{tabular}{ccccccccccccc}
\hline
&\multicolumn{4}{c}{$^{92}$Mo/$^{94}$Mo}&\multicolumn{4}{c}{$^{84}$Sr/$^{94}$Mo}&\multicolumn{4}{c}{$^{78}$Kr/$^{94}$Mo}\\
Trajectory&$\left.\frac{Y(92)}{Y(94)}\right|_\mathrm{peak}$&$\left.\frac{Y(92)}{Y(94)} 
\right|_\mathrm{50}$&Up&Down&$\left.\frac{Y(84)}{Y(94)}\right|_\mathrm{peak}$&$\left.\frac{Y(84)}{Y(94)} 
\right|_\mathrm{50}$&Up&Down&$\left.\frac{Y(78)}{Y(94)}\right|_\mathrm{peak}$&$\left.\frac{Y(78)}{Y(94)} 
\right|_\mathrm{50}$&Up&Down\\
\hline
\#06&2.00 &2.60 &2.24&0.770&76.5 &99.4 &5.49&0.336&194   &718  &24.8&0.608\\
\#11&0.923&1.20 &2.14&0.793&1.86 &2.41 &3.03&0.627&2.18  &2.83 &3.64&0.547\\
\#16&0.631&0.820&2.79&0.666&0.213&0.277&2.50&0.618&0.0837&0.109&2.76&0.573\\
\#19&0.876&1.14 &2.98&0.627&0.530&0.689&2.37&0.611&0.311 &0.404&2.47&0.618\\
\#21&0.980&1.27 &2.87&0.675&0.664&0.862&2.25&0.744&0.390 &0.507&2.32&0.733\\
\#23&0.983&1.28 &2.85&0.651&0.693&0.900&2.23&0.766&0.393 &0.511&2.32&0.749\\
\hline
\end{tabular}
\end{table*}

\section{Summary and conclusions}
\label{sec:summary}

A comprehensive, large-scale MC study of nucleosynthesis in the \nup process has been performed. A range of conditions in a $\ye$ and entropy parameter-space was explored to cover the possibilities regarding implementations of a \nup process in different sites. Our results allow to quantify the uncertainties stemming from nuclear physics input for any particular astrophysical simulation spanning this wide range of $\ye$ and entropy parameter-space.

For each of 23 chosen trajectories, and a choice for the \triplea and $\iso{Ni}{56}$(n,\,p)$\iso{Co}{56}$ reaction rates, the astrophysical reaction rates for several thousand target nuclides for Fe and above were simultaneously varied within individual temperature-dependent uncertainty ranges constructed from a combination of experimental and theoretical error bars. This allowed the investigation of the combined effect of rate uncertainties leading to total uncertainties in the final abundances of stable nuclei obtained after the \nup process had ceased. Key rates dominating the uncertainties in the final yields were determined. Different key rates were found for each trajectory as the production range of nuclides depends on the thermodynamic conditions.

The rates for the \triplea and the $\iso{Ni}{56}({\rm n},{\rm p})\iso{Co}{56}$ reaction were not included in the MC variation because their uncertainties dominate the production uncertainties of all nuclides and therefore would cover any other key rates. They should be considered as key reactions, nevertheless.

Among the other key reactions found, $({\rm n},{\rm p})$ reactions dominate because they determine the flow from one isotonic chain into the next. Most proton captures are in equilibrium and therefore their individual rates are not important. Several (p,\,$\gamma$) rates having been identified as key rates are at the edge of the reaction flow or fall out of equilibrium within our variation limits. Among those is the proton capture on the stable nuclide \iso{Mo}{92}, acting at late times and affecting the abundance of the $p$ nuclide \iso{Mo}{92}, provided the conditions of trajectories \#16 or \#17 are found in nature. Similarly, the reaction {\iso{Ru}{94}}(p,$\gamma$)\iso{Rh}{95} is a key reaction for the $p$ nucleus \iso{Mo}{94}.

Concerning the isotope ratios of light $p$ nuclides it was found that it is possible to reproduce the solar \iso{Mo}{92}/\iso{Mo}{94} abundance ratio within uncertainties, even though only rate uncertainties and not mass uncertainties have been considered. The reproduction of both the Mo isotopic ratio and their production level relative to the lighter $p$ isotopes of Kr and Sr has been found to be difficult within one trajectory. It has to be cautioned, however, that a contribution to the Mo isotopes stemming from the proton-rich side is severely constrained by the fact that live \iso{Nb}{92} was found in the early solar system, which cannot be produced by the decay of proton-rich, unstable progenitor nuclei \citep{2003NuPhA.719..287D,2019arXiv190507828C}. It has to be noted further that realistic sites may give rise to a range of conditions, resembling a combination of several of our trajectories with different weights. The range of conditions and their respective weights may also depend on the specific nucleosynthesis site and may be different for different sites. A parameter study like the present investigation is not devised to address such a superposition of conditions. Once site conditions have been constrained by hydrodynamical studies, however, our results can be used to assess the feasibility to reproduce abundance patterns of the solar system and those found in meteorites. Therefore, for the time being -- before having further constrained nucleosynthesis sites and reaction rates -- it has to be concluded that a consistent production of the light $p$ nuclides (including the Mo isotopes) in the \nup process cannot be ruled out. We also can conclude that uncertainties in nuclear reaction rates may still have equal or even stronger impact than mass uncertainties in the path of the \nup process.

In summary, we found that the uncertainties in the production of nuclei are dominated by the uncertainties arising from the choice of site, explosion model, and numerical treatment of the explosion hydrodynamics, as these crucially determine what range of nuclei can actually be produced. Although the astrophysical constraints seem to be similarly weak for the $\nu p$ process as for the $r$ process, the \nup process is better constrained by nuclear physics and exhibits smaller uncertainties therein, at least in the dominating rates. Uncertainties stemming from the astrophysical reaction rates become important only after the nucleosynthesis conditions have been constrained better. Nevertheless, an experimental verification of the predicted rates will be difficult, not only because of the short-lived, intermediate, and heavy nuclei involved but also due to the high plasma temperatures, giving rise to considerable thermal excitation and thus small ground state contributions to the stellar rate \citep{2012ApJS..201...26R, rauadvance}. Importantly, even where feasible, experimental cross section data typically only constrain these ground-state contributions. More promising is the experimental determination of nuclear properties required for the calculation of nuclear reaction rates. These not only include masses but, more importantly, also excitation energies, spins, and parities of excited states, both below the proton separation energy and in the relevant Gamow window. The determination of particle widths would improve constraints on the key reactions involving protons and $\alpha$ particles. Present and future facilities using unstable beams offer possibilities for extracting such information.

\section*{Acknowledgements}

The authors thank the reviewer for her/his valuable comments on the submitted manuscript. N.N. thanks T. Fischer for providing neutrino-driven wind trajectories used in the present study. This work has been partially supported by the European Research Council (EU-FP7-ERC-2012-St Grant 306901, EU-FP7 Adv Grant GA321263-FISH), the EU COST Action CA16117 (ChETEC), the UK STFC (ST/M000958/1), and MEXT Japan (``Priority Issue on Post-K computer: Elucidation of the Fundamental Laws and Evolution of the Universe'' and ``the World Premier International Research Centre Initiative: WPI Initiative''). G.C. acknowledges financial support from the EU Horizon2020 programme under the Marie Sk\l odowska-Curie grant 664931. C.F. acknowledges support by the U.S. Department of Energy, Office of Science, Office of Nuclear Physics, under Award No.~DE-FG02-02ER41216. Parts of the computations were carried out on COSMOS (STFC DiRAC Facility) at DAMTP in University of Cambridge. This equipment was funded by BIS National E-infrastructure capital grant ST/J005673/1, STFC capital grant ST/H008586/1, and STFC DiRAC Operations grant ST/K00333X/1. DiRAC is part of the UK National E-Infrastructure. Further computations were carried out at CfCA, National Astronomical Observatory of Japan, and at YITP, Kyoto University. The University of Edinburgh is a charitable body, registered in Scotland, with Registration No.~SC005336.




\bibliographystyle{mnras}
\bibliography{ref} 

\begin{thebibliography}{}
\makeatletter
\relax
\def\mn@urlcharsother{\let\do\@makeother \do\$\do\&\do\#\do\^\do\_\do\%\do\~}
\def\mn@doi{\begingroup\mn@urlcharsother \@ifnextchar [ {\mn@doi@}
  {\mn@doi@[]}}
\def\mn@doi@[#1]#2{\def\@tempa{#1}\ifx\@tempa\@empty \href
  {http://dx.doi.org/#2} {doi:#2}\else \href {http://dx.doi.org/#2} {#1}\fi
  \endgroup}
\def\mn@eprint#1#2{\mn@eprint@#1:#2::\@nil}
\def\mn@eprint@arXiv#1{\href {http://arxiv.org/abs/#1} {{\tt arXiv:#1}}}
\def\mn@eprint@dblp#1{\href {http://dblp.uni-trier.de/rec/bibtex/#1.xml}
  {dblp:#1}}
\def\mn@eprint@#1:#2:#3:#4\@nil{\def\@tempa {#1}\def\@tempb {#2}\def\@tempc
  {#3}\ifx \@tempc \@empty \let \@tempc \@tempb \let \@tempb \@tempa \fi \ifx
  \@tempb \@empty \def\@tempb {arXiv}\fi \@ifundefined
  {mn@eprint@\@tempb}{\@tempb:\@tempc}{\expandafter \expandafter \csname
  mn@eprint@\@tempb\endcsname \expandafter{\@tempc}}}

\bibitem[\protect\citeauthoryear{{Aikawa}, {Arnould}, {Goriely}, {Jorissen}  \&
  {Takahashi}}{{Aikawa} et~al.}{2005}]{2005A&A...441.1195A}
{Aikawa} M.,  {Arnould} M.,  {Goriely} S.,  {Jorissen} A.,   {Takahashi} K.,
  2005, \mn@doi [\aap] {10.1051/0004-6361:20052944}, \href
  {http://adsabs.harvard.edu/abs/2005A%26A...441.1195A} {441, 1195}

\bibitem[\protect\citeauthoryear{{Angulo} et~al.,}{{Angulo}
  et~al.}{1999}]{1999NuPhA.656....3A}
{Angulo} C.,  et~al., 1999, \mn@doi [Nuclear Physics A]
  {10.1016/S0375-9474(99)00030-5}, \href
  {http://adsabs.harvard.edu/abs/1999NuPhA.656....3A} {656, 3}

\bibitem[\protect\citeauthoryear{{Arcones} \& {Bliss}}{{Arcones} \&
  {Bliss}}{2014}]{2014JPhG...41d4005A}
{Arcones} A.,  {Bliss} J.,  2014, \mn@doi [Journal of Physics G Nuclear
  Physics] {10.1088/0954-3899/41/4/044005}, \href
  {http://adsabs.harvard.edu/abs/2014JPhG...41d4005A} {41, 044005}

\bibitem[\protect\citeauthoryear{{Arcones}, {Fr{\"o}hlich}  \&
  {Mart{\'{\i}}nez-Pinedo}}{{Arcones} et~al.}{2012}]{arcofromart}
{Arcones} A.,  {Fr{\"o}hlich} C.,   {Mart{\'{\i}}nez-Pinedo} G.,  2012, \mn@doi
  [\apj] {10.1088/0004-637X/750/1/18}, \href
  {http://adsabs.harvard.edu/abs/2012ApJ...750...18A} {750, 18}

\bibitem[\protect\citeauthoryear{{Bliss}, {Witt}, {Arcones}, {Montes}  \&
  {Pereira}}{{Bliss} et~al.}{2018a}]{2018ApJ...855..135B}
{Bliss} J.,  {Witt} M.,  {Arcones} A.,  {Montes} F.,   {Pereira} J.,  2018a,
  \mn@doi [\apj] {10.3847/1538-4357/aaadbe}, \href
  {http://adsabs.harvard.edu/abs/2018ApJ...855..135B} {855, 135}

\bibitem[\protect\citeauthoryear{{Bliss}, {Arcones}  \& {Qian}}{{Bliss}
  et~al.}{2018b}]{2018ApJ...866..105B}
{Bliss} J.,  {Arcones} A.,   {Qian} Y.-Z.,  2018b, \mn@doi [\apj]
  {10.3847/1538-4357/aade8d}, \href
  {http://adsabs.harvard.edu/abs/2018ApJ...866..105B} {866, 105}

\bibitem[\protect\citeauthoryear{{Caughlan} \& {Fowler}}{{Caughlan} \&
  {Fowler}}{1988}]{1988ADNDT..40..283C}
{Caughlan} G.~R.,  {Fowler} W.~A.,  1988, \mn@doi [Atomic Data and Nuclear Data
  Tables] {10.1016/0092-640X(88)90009-5}, \href
  {http://adsabs.harvard.edu/abs/1988ADNDT..40..283C} {40, 283}

\bibitem[\protect\citeauthoryear{{Cescutti}, {Hirschi}, {Nishimura}, {Hartogh},
  {Rauscher}, {Murphy}  \& {Cristallo}}{{Cescutti}
  et~al.}{2018}]{2018MNRAS.478.4101C}
{Cescutti} G.,  {Hirschi} R.,  {Nishimura} N.,  {Hartogh} J.~W.~d.,  {Rauscher}
  T.,  {Murphy} A.~S.~J.,   {Cristallo} S.,  2018, \mn@doi [\mnras]
  {10.1093/mnras/sty1185}, \href
  {http://adsabs.harvard.edu/abs/2018MNRAS.478.4101C} {478, 4101}

\bibitem[\protect\citeauthoryear{{C{\^o}t{\'e}}, {Lugaro}, {Reifarth},
  {Pignatari}, {Vil{\'a}gos}, {Yag{\"u}e}  \& {Gibson}}{{C{\^o}t{\'e}}
  et~al.}{2019}]{2019arXiv190507828C}
{C{\^o}t{\'e}} B.,  {Lugaro} M.,  {Reifarth} R.,  {Pignatari} M.,
  {Vil{\'a}gos} B.,  {Yag{\"u}e} A.,   {Gibson} B.~K.,  2019, \apj, \href
  {https://ui.adsabs.harvard.edu/abs/2019arXiv190507828C} {in press;
  arXiv:1905.07828}

\bibitem[\protect\citeauthoryear{{Cyburt} et~al.,}{{Cyburt}
  et~al.}{2010}]{2010ApJS..189..240C}
{Cyburt} R.~H.,  et~al., 2010, \mn@doi [\apjs] {10.1088/0067-0049/189/1/240},
  \href {http://adsabs.harvard.edu/abs/2010ApJS..189..240C} {189, 240}

\bibitem[\protect\citeauthoryear{{Dauphas}, {Rauscher}, {Marty}  \&
  {Reisberg}}{{Dauphas} et~al.}{2003}]{2003NuPhA.719..287D}
{Dauphas} N.,  {Rauscher} T.,  {Marty} B.,   {Reisberg} L.,  2003, \mn@doi
  [Nuclear Physics A] {10.1016/S0375-9474(03)00934-5}, \href
  {https://ui.adsabs.harvard.edu/abs/2003NuPhA.719..287D} {719, C287}

\bibitem[\protect\citeauthoryear{{Dillmann}, {Heil}, {K{\"a}ppeler}, {Plag},
  {Rauscher}  \& {Thielemann}}{{Dillmann} et~al.}{2006}]{2006AIPC..819..123D}
{Dillmann} I.,  {Heil} M.,  {K{\"a}ppeler} F.,  {Plag} R.,  {Rauscher} T.,
  {Thielemann} F.-K.,  2006, in {Woehr} A.,  {Aprahamian} A.,  eds,  American
  Institute of Physics Conference Series Vol. 819, Capture Gamma-Ray
  Spectroscopy and Related Topics. pp 123--127, \mn@doi{10.1063/1.2187846}

\bibitem[\protect\citeauthoryear{{Fisker}, {Hoffman}  \& {Pruet}}{{Fisker}
  et~al.}{2009}]{2009ApJ...690L.135F}
{Fisker} J.~L.,  {Hoffman} R.~D.,   {Pruet} J.,  2009, \mn@doi [\apjl]
  {10.1088/0004-637X/690/2/L135}, \href
  {https://ui.adsabs.harvard.edu/abs/2009ApJ...690L.135F} {690, L135}

\bibitem[\protect\citeauthoryear{{Fran{\c c}ois} et~al.,}{{Fran{\c c}ois}
  et~al.}{2007}]{2007A&A...476..935F}
{Fran{\c c}ois} P.,  et~al., 2007, \mn@doi [\aap] {10.1051/0004-6361:20077706},
  \href {http://adsabs.harvard.edu/abs/2007A%26A...476..935F} {476, 935}

\bibitem[\protect\citeauthoryear{{Freiburghaus} \& {Rauscher}}{{Freiburghaus}
  \& {Rauscher}}{1999}]{freiburghaus1999}
{Freiburghaus} C.,  {Rauscher} T.,  1999, Reaction rate library in REACLIB
  format, available at: http://nucastro.org/reaclib

\bibitem[\protect\citeauthoryear{{Fr{\"o}hlich} \& {Hatcher}}{{Fr{\"o}hlich} \&
  {Hatcher}}{2015}]{2015EPJWC..9303008F}
{Fr{\"o}hlich} C.,  {Hatcher} D.,  2015, in European Physical Journal Web of
  Conferences. p. 03008, \mn@doi{10.1051/epjconf/20159303008}

\bibitem[\protect\citeauthoryear{Fr\"ohlich, Mart\'inez-Pinedo, Liebend\"orfer,
  Thielemann, Bravo, Hix, Langanke  \& Zinner}{Fr\"ohlich
  et~al.}{2006a}]{fro06a}
Fr\"ohlich C.,  Mart\'inez-Pinedo G.,  Liebend\"orfer M.,  Thielemann F.-K.,
  Bravo E.,  Hix W.~R.,  Langanke K.,   Zinner N.~T.,  2006a, \prl, 96, 142502

\bibitem[\protect\citeauthoryear{Fr\"ohlich et~al.,}{Fr\"ohlich
  et~al.}{2006b}]{fro06b}
Fr\"ohlich C.,  et~al., 2006b, \apj, 637, 415

\bibitem[\protect\citeauthoryear{{Fujibayashi}, {Yoshida}  \&
  {Sekiguchi}}{{Fujibayashi} et~al.}{2015}]{2015ApJ...810..115F}
{Fujibayashi} S.,  {Yoshida} T.,   {Sekiguchi} Y.,  2015, \mn@doi [\apj]
  {10.1088/0004-637X/810/2/115}, \href
  {http://adsabs.harvard.edu/abs/2015ApJ...810..115F} {810, 115}

\bibitem[\protect\citeauthoryear{{Fynbo} et~al.,}{{Fynbo}
  et~al.}{2005}]{2005Natur.433..136F}
{Fynbo} H.~O.~U.,  et~al., 2005, \nat, \href
  {http://adsabs.harvard.edu/abs/2005Natur.433..136F} {433, 136}

\bibitem[\protect\citeauthoryear{{Goriely}}{{Goriely}}{1999}]{1999A&A...342..881G}
{Goriely} S.,  1999, \aap, \href
  {http://adsabs.harvard.edu/abs/1999A%26A...342..881G} {342, 881}

\bibitem[\protect\citeauthoryear{Haettner et~al.,}{Haettner
  et~al.}{2011}]{PhysRevLett.106.122501}
Haettner E.,  et~al., 2011, \mn@doi [Phys. Rev. Lett.]
  {10.1103/PhysRevLett.106.122501}, 106, 122501

\bibitem[\protect\citeauthoryear{{Kizivat}, {Mart{\'{\i}}nez-Pinedo},
  {Langanke}, {Surman}  \& {McLaughlin}}{{Kizivat}
  et~al.}{2010}]{2010PhRvC..81b5802K}
{Kizivat} L.-T.,  {Mart{\'{\i}}nez-Pinedo} G.,  {Langanke} K.,  {Surman} R.,
  {McLaughlin} G.~C.,  2010, \mn@doi [\prc] {10.1103/PhysRevC.81.025802}, \href
  {http://adsabs.harvard.edu/abs/2010PhRvC..81b5802K} {81, 025802}

\bibitem[\protect\citeauthoryear{{Lodders}}{{Lodders}}{2003}]{2003ApJ...591.1220L}
{Lodders} K.,  2003, \mn@doi [\apj] {10.1086/375492}, \href
  {https://ui.adsabs.harvard.edu/abs/2003ApJ...591.1220L} {591, 1220}

\bibitem[\protect\citeauthoryear{{Mayer}, {Goriely}, {Netterdon}, {P{\'e}ru},
  {Scholz}, {Schwengner}  \& {Zilges}}{{Mayer}
  et~al.}{2016}]{2016PhRvC..93d5809M}
{Mayer} J.,  {Goriely} S.,  {Netterdon} L.,  {P{\'e}ru} S.,  {Scholz} P.,
  {Schwengner} R.,   {Zilges} A.,  2016, \mn@doi [\prc]
  {10.1103/PhysRevC.93.045809}, \href
  {https://ui.adsabs.harvard.edu/abs/2016PhRvC..93d5809M} {93, 045809}

\bibitem[\protect\citeauthoryear{{Montes} et~al.,}{{Montes}
  et~al.}{2007}]{2007ApJ...671.1685M}
{Montes} F.,  et~al., 2007, \mn@doi [\apj] {10.1086/523084}, \href
  {http://adsabs.harvard.edu/abs/2007ApJ...671.1685M} {671, 1685}

\bibitem[\protect\citeauthoryear{{Nishimura} et~al.,}{{Nishimura}
  et~al.}{2012}]{2012ApJ...758....9N}
{Nishimura} N.,  et~al., 2012, \mn@doi [\apj] {10.1088/0004-637X/758/1/9},
  \href {http://adsabs.harvard.edu/abs/2012ApJ...758....9N} {758, 9}

\bibitem[\protect\citeauthoryear{{Nishimura}, {Hirschi}, {Rauscher},
  {St.~J.~Murphy}  \& {Cescutti}}{{Nishimura}
  et~al.}{2017}]{2017MNRAS.469.1752N}
{Nishimura} N.,  {Hirschi} R.,  {Rauscher} T.,  {St.~J.~Murphy} A.,
  {Cescutti} G.,  2017, \mn@doi [\mnras] {10.1093/mnras/stx696}, \href
  {http://adsabs.harvard.edu/abs/2017MNRAS.469.1752N} {469, 1752}

\bibitem[\protect\citeauthoryear{{Nishimura}, {Rauscher}, {Hirschi}, {Murphy},
  {Cescutti}  \& {Travaglio}}{{Nishimura} et~al.}{2018}]{2018MNRAS.474.3133N}
{Nishimura} N.,  {Rauscher} T.,  {Hirschi} R.,  {Murphy} A.~S.~J.,  {Cescutti}
  G.,   {Travaglio} C.,  2018, \mn@doi [\mnras] {10.1093/mnras/stx3033}, \href
  {http://adsabs.harvard.edu/abs/2018MNRAS.474.3133N} {474, 3133}

\bibitem[\protect\citeauthoryear{Pearson \& Galton}{Pearson \&
  Galton}{1895}]{doi:10.1098/rspl.1895.0041}
Pearson K.,  Galton F.,  1895, \mn@doi [Proceedings of the Royal Society of
  London] {10.1098/rspl.1895.0041}, 58, 240

\bibitem[\protect\citeauthoryear{{Pruet}, {Hoffman}, {Woosley}, {Janka}  \&
  {Buras}}{{Pruet} et~al.}{2006}]{pruet06}
{Pruet} J.,  {Hoffman} R.~D.,  {Woosley} S.~E.,  {Janka} H.-T.,   {Buras} R.,
  2006, \mn@doi [\apj] {10.1086/503891}, \href
  {http://ads.nao.ac.jp/abs/2006ApJ...644.1028P} {644, 1028}

\bibitem[\protect\citeauthoryear{{Rauscher}}{{Rauscher}}{2012}]{2012ApJS..201...26R}
{Rauscher} T.,  2012, \mn@doi [\apjs] {10.1088/0067-0049/201/2/26}, \href
  {http://adsabs.harvard.edu/abs/2012ApJS..201...26R} {201, 26}

\bibitem[\protect\citeauthoryear{{Rauscher}}{{Rauscher}}{2014}]{rauadvance}
{Rauscher} T.,  2014, \mn@doi [AIP Advances] {10.1063/1.4868239}, \href
  {http://adsabs.harvard.edu/abs/2014AIPA....4d1012R} {4, 041012}

\bibitem[\protect\citeauthoryear{{Rauscher} \& {Thielemann}}{{Rauscher} \&
  {Thielemann}}{2000}]{2000ADNDT..75....1R}
{Rauscher} T.,  {Thielemann} F.-K.,  2000, \mn@doi [Atomic Data and Nuclear
  Data Tables] {10.1006/adnd.2000.0834}, \href
  {http://adsabs.harvard.edu/abs/2000ADNDT..75....1R} {75, 1}

\bibitem[\protect\citeauthoryear{{Rauscher}, {Dauphas}, {Dillmann},
  {Fr{\"o}hlich}, {F{\"u}l{\"o}p}  \& {Gy{\"u}rky}}{{Rauscher}
  et~al.}{2013}]{2013RPPh...76f6201R}
{Rauscher} T.,  {Dauphas} N.,  {Dillmann} I.,  {Fr{\"o}hlich} C.,
  {F{\"u}l{\"o}p} Z.,   {Gy{\"u}rky} G.,  2013, \mn@doi [Reports on Progress in
  Physics] {10.1088/0034-4885/76/6/066201}, \href
  {http://adsabs.harvard.edu/abs/2013RPPh...76f6201R} {76, 066201}

\bibitem[\protect\citeauthoryear{{Rauscher}, {Nishimura}, {Hirschi},
  {Cescutti}, {Murphy}  \& {Heger}}{{Rauscher}
  et~al.}{2016}]{2016MNRAS.463.4153R}
{Rauscher} T.,  {Nishimura} N.,  {Hirschi} R.,  {Cescutti} G.,  {Murphy}
  A.~S.~J.,   {Heger} A.,  2016, \mn@doi [\mnras] {10.1093/mnras/stw2266},
  \href {http://adsabs.harvard.edu/abs/2016MNRAS.463.4153R} {463, 4153}

\bibitem[\protect\citeauthoryear{{Rauscher}, {Nishimura}, {Cescutti}, {Hirschi}
   \& {Murphy}}{{Rauscher} et~al.}{2018}]{2018AIPC.1947b0015R}
{Rauscher} T.,  {Nishimura} N.,  {Cescutti} G.,  {Hirschi} R.,   {Murphy}
  A.~S.~J.,  2018, in American Institute of Physics Conference Series. p.
  020015 (\mn@eprint {arXiv} {1709.00690}), \mn@doi{10.1063/1.5030819}

\bibitem[\protect\citeauthoryear{{Sallaska}, {Iliadis}, {Champange}, {Goriely},
  {Starrfield}  \& {Timmes}}{{Sallaska} et~al.}{2013}]{2013ApJS..207...18S}
{Sallaska} A.~L.,  {Iliadis} C.,  {Champange} A.~E.,  {Goriely} S.,
  {Starrfield} S.,   {Timmes} F.~X.,  2013, \mn@doi [\apjs]
  {10.1088/0067-0049/207/1/18}, \href
  {http://adsabs.harvard.edu/abs/2013ApJS..207...18S} {207, 18}

\bibitem[\protect\citeauthoryear{{Schatz} et~al.,}{{Schatz}
  et~al.}{1998}]{schatz}
{Schatz} H.,  et~al., 1998, \mn@doi [\physrep] {10.1016/S0370-1573(97)00048-3},
  \href {http://adsabs.harvard.edu/abs/1998PhR...294..167S} {294, 167}

\bibitem[\protect\citeauthoryear{{Takahashi} \& {Yokoi}}{{Takahashi} \&
  {Yokoi}}{1987}]{1987ADNDT..36..375T}
{Takahashi} K.,  {Yokoi} K.,  1987, \mn@doi [Atomic Data and Nuclear Data
  Tables] {10.1016/0092-640X(87)90010-6}, \href
  {http://adsabs.harvard.edu/abs/1987ADNDT..36..375T} {36, 375}

\bibitem[\protect\citeauthoryear{{Wanajo}}{{Wanajo}}{2006}]{2006ApJ...647.1323W}
{Wanajo} S.,  2006, \mn@doi [\apj] {10.1086/505483}, \href
  {http://ads.nao.ac.jp/abs/2006ApJ...647.1323W} {647, 1323}

\bibitem[\protect\citeauthoryear{{Wanajo}, {Janka}  \& {Kubono}}{{Wanajo}
  et~al.}{2011}]{2011ApJ...729...46W}
{Wanajo} S.,  {Janka} H.-T.,   {Kubono} S.,  2011, \mn@doi [\apj]
  {10.1088/0004-637X/729/1/46}, \href
  {http://adsabs.harvard.edu/abs/2011ApJ...729...46W} {729, 46}

\bibitem[\protect\citeauthoryear{Weber et~al.,}{Weber
  et~al.}{2008}]{PhysRevC.78.054310}
Weber C.,  et~al., 2008, \mn@doi [Phys. Rev. C] {10.1103/PhysRevC.78.054310},
  78, 054310

\bibitem[\protect\citeauthoryear{{Woosley} et~al.,}{{Woosley}
  et~al.}{2004}]{2004ApJS..151...75W}
{Woosley} S.~E.,  et~al., 2004, \mn@doi [\apjs] {10.1086/381533}, \href
  {https://ui.adsabs.harvard.edu/abs/2004ApJS..151...75W} {151, 75}

\bibitem[\protect\citeauthoryear{Xing et~al.,}{Xing et~al.}{2018}]{XING2018358}
Xing Y.,  et~al., 2018, \mn@doi [Physics Letters B]
  {https://doi.org/10.1016/j.physletb.2018.04.009}, 781, 358

\bibitem[\protect\citeauthoryear{{Xu}, {Takahashi}, {Goriely}, {Arnould},
  {Ohta}  \& {Utsunomiya}}{{Xu} et~al.}{2013}]{2013NuPhA.918...61X}
{Xu} Y.,  {Takahashi} K.,  {Goriely} S.,  {Arnould} M.,  {Ohta} M.,
  {Utsunomiya} H.,  2013, \mn@doi [Nuclear Physics A]
  {10.1016/j.nuclphysa.2013.09.007}, \href
  {http://adsabs.harvard.edu/abs/2013NuPhA.918...61X} {918, 61}

\makeatother
\end{thebibliography}



%
%


\bsp	
\label{lastpage}
\end{document}